\documentclass[floatfix,%
 aip,
 pop,
 amsmath,amssymb,
 reprint,
]{revtex4-2}
\usepackage{multirow}
\usepackage{graphicx}
\usepackage{xcolor}
\usepackage{dcolumn}
\usepackage{bm}
\usepackage[utf8]{inputenc}
\usepackage[T1]{fontenc}
\usepackage{mathptmx}
\usepackage{soul}
\usepackage{array}
\usepackage{comment}

\begin{document}

\title{Experimental observation of higher-harmonic generation of whistler waves in magnetized plasma of Large Volume Plasma Device }

\author{Deepak Rajpurohit}
\email{draj91297@gmail.com}
\affiliation{Institute for Plasma Research, Gandhinagar 382428, India}
\affiliation{Homi Bhabha National Institute, Training School Complex, Anushaktinagar, Mumbai 400094, India}

\author{A. K. Sanyasi}
\email{amulya@ipr.res.in}
\affiliation{Institute for Plasma Research, Gandhinagar 382428, India}
\affiliation{Homi Bhabha National Institute, Training School Complex, Anushaktinagar, Mumbai 400094, India}

\author{P. K. Srivastava}
\affiliation{Institute for Plasma Research, Gandhinagar 382428, India}

\author{Ayan Adhikari}
\affiliation{Institute for Plasma Research, Gandhinagar 382428, India}
\affiliation{Homi Bhabha National Institute, Training School Complex, Anushaktinagar, Mumbai 400094, India}

\date{\today}


\begin{abstract}
We report the higher-harmonic generation of the whistler waves using a loop antenna in the LVPD. The antenna is energized using a pulse-forming network (PFN), and the excited magnetic fluctuations are measured using a three-axis B-dot probe. Experiments are performed in a moderate-density ($n_e\sim1$--$2\times10^{17},\mathrm{m}^{-3}$), low-temperature ($T_e\sim2$--$5,\mathrm{eV}$) argon plasma in a uniform axial magnetic field ($B_z\sim6.2,\mathrm{G}$). A Helmholtz coil is employed to investigate the dependence of the wave response and propagation on the local magnetic-field configuration.
A clear distinction between wave characteristics is observed in the absence and presence of plasma: in plasma, the excited wave shows a significant $\delta B_y$ component along with $\delta B_x$, indicating polarization. In addition to the fundamental frequency, distinct spectral peaks are observed at $300\,\mathrm{kHz}$, $450\,\mathrm{kHz}$, and $600\,\mathrm{kHz}$. These harmonics indicate nonlinear wave--plasma interactions. Bicoherence analysis further provides evidence of nonlinear coupling between the observed frequency components. The spatial propagation of the wave was systematically diagnosed using a linear array of five identical B-dot probes. Spatial cross-correlation analysis reveals strong structural coherence during wave propagation, allowing the explicit experimental determination of the wave phase velocity ($V_{ph}$) and the parallel wave number ($k_{\parallel}$) for the higher harmonics generated. The measured propagation is oblique, with a propagation angle of approximately $48^\circ$ with respect to the background magnetic field. The experimentally determined dispersion relation is compared with the cold-plasma oblique whistler dispersion relation and is found to be consistent within experimental uncertainty. The study contributes to the fundamental understanding of nonlinear wave dynamics in laboratory magnetized plasmas.
\end{abstract}

\maketitle

\section{Introduction}

 \hspace{0.25cm}
The study of electromagnetic (EM) wave propagation~\cite{35} and interactions with plasma has received significant attention due to its wide range of applications in telecommunication, space explorations, and controlled nuclear fusion~\cite{13,25,34}. EM waves can be self-generated or excited via different mechanisms. In particular, using a straight wire or a circular loop antenna~\cite{28,53}, EM waves can be excited in vacuum and plasma environments, as well as in the presence of a guiding magnetic field. This presents a compelling method for exploring wave--plasma interaction studies~\cite{34}. Electromagnetic waves in the whistler-frequency range are naturally observed in the Earth's magnetosphere~\cite{36,37,38,39}. These waves are mostly excited in magnetized plasmas by Antenna\cite{55}, energetic electrons, electron beams, electron-temperature anisotropy\cite{56}, loss--cone reflected electrons, etc.~\cite{12,18,4}. Recent laboratory experiments have successfully reproduced key excitation mechanisms of naturally occurring whistler-mode waves observed in near-Earth space. Van Compernolle et al.\cite{55} demonstrated the generation of chorus-like chirping whistler emissions through the injection of an energetic oblique electron beam into a magnetized plasma, reproducing rising-tone, falling-tone, and hiss-like emissions analogous to those observed in the Earth's radiation belts. Ma et al.\cite{56} reported the controlled excitation of whistler waves driven by electron temperature anisotropy produced through electron cyclotron resonance heating in a magnetic mirror configuration. These laboratory investigations provide direct experimental validation of mechanisms believed to operate in the terrestrial magnetosphere, where anisotropic and energetic electron populations drive whistler-mode chorus emissions responsible for radiation-belt electron acceleration, precipitation, and broader space-weather dynamics. An et al.~\cite{58} experimentally investigated the resonant excitation of whistler waves in a laboratory plasma and reconstructed the wave mode structure using phase-correlation measurements. Their study experimentally identified the wave propagation characteristics and resonance mechanisms responsible for whistler-wave excitation.
Whistlers propagate in the frequency range $\omega_{ci} \ll \omega_{lh} \approx \omega_{wh} < \omega_{ce}$, with wave propagation $(k_{\parallel})$ phase velocity close to its resonance cone having its maxima at $\omega_{wh} = 0.5\omega_{ce}$~\cite{4}, where $\omega_{ci}$, $\omega_{lh}$ ($\omega_{lh}=\sqrt{\omega_{ci}\,\omega_{ce}}$), $\omega_{ce}$, and $\omega_{wh}$ are the ion-cyclotron, lower-hybrid, electron-cyclotron, and whistler-wave frequencies, respectively.

Higher-Harmonic Generation (HHG) in electromagnetic waves is of fundamental interest in the field of wave–plasma interactions ~\cite{5,6,7,8}. It is frequently observed in laser–plasma interactions~\cite{31,32} and electron plasma waves, where nonlinear wave–plasma coupling leads to the generation of harmonics of the fundamental wave. The theoretical investigation of nonlinear interactions between electromagnetic waves and a plasma layer in a magnetized plasma was carried out by Whitmer et al.~\cite{49,50,51,52}. Subsequently, Krenz and Kino~\cite{3} developed the theoretical framework for parametric oscillations and harmonic generation in rf- plasma.
Furthermore, Kaw and Sodha~\cite{5,6,7,8} made seminal contributions through a series of theoretical papers, significantly advancing the understanding of wave harmonic generation in plasma media. However, similar phenomena in whistler-mode and other ionospheric plasma waves remain relatively unexplored. One of the first observational efforts in this domain was led by Birbal Singh~\cite{15}, who, for the first time, observed and recorded higher harmonic adjustments in the ionosphere from a ground-based low-latitude station. Subsequently, in 2015, Tsugawa et al.~\cite{16} reported the presence of higher harmonics of whistler waves in the near-Moon region, detected by the Kaguya spacecraft. It was reported that the possible cause for higher-harmonic generation is due to the intrinsic environment of the Moon, such as lunar dust and local magnetic anomalies that could have promoted the nonlinear steepening of waveforms. Recently, in 2017, Punit Kumar et al.~\cite{29} presented a theoretical and analytical explanation for the second harmonic generation of whistler pulses in quantum magnetized plasma using the quantum hydrodynamic (QHD) model, and in 2021, Xiongdong Yu et al.~\cite{17} investigated the mechanisms of HHG in electromagnetic waves within magnetized plasmas, employing multiscale analysis~\cite{40} and a kinetic approach~\cite{13,41}.The generation of higher harmonics was observed in laboratory plasma~\cite{48} in 1998 but exact physical phenomena behind, conversion efficiency, and spatial variations of higher-harmonic generation of whistler waves in laboratory plasma remains an open question. Therefore, it is crucial to design and conduct laboratory experiments to verify and understand the mechanisms behind HHG in whistler-mode waves.
In the present study, electromagnetic waves were launched in the Large Volume Plasma Device-Upgrade (LVPD-U) using a circular loop antenna under systematically varied experimental conditions. The experimental conditions are described as, (a) in vacuum, under different magnetic field configurations, to characterize the antenna response; (b) in plasma, within a uniform axial magnetic field ($B_z = 6.2\text{ G}$, $B_H = 0\text{ G}$) at varying exciter currents ($6\text{ A}$, $10\text{ A}$, and $50\text{ A}$), to evaluate the linear and nonlinear wave response; and (c) in the presence of superposed Helmholtz fields ($B_H = \pm 20\text{ G}$, $B_z=6.2G$) at a higher driving current ($50\text{ A}$), to investigate the nonlinear wave response under different magnetic field configurations.

Section~2 describes the experimental setup, plasma characterization is discussed in Section~3, Section~4 elaborates on the electromagnetic-wave excitation experiments, and Section~5 provides experimental investigations into wave mode characterization. The mathematical formalism and mechanisms of HHG are briefly discussed in Section~6, and observations and results are concluded in Section~7. 
\section{Experimental Setup and Device Configuration}
\subsection{Large Volume Plasma Device-Upgrade (LVPD-U)}
\begin{figure*}[t]
    \centering
    \includegraphics[width=0.95\textwidth]{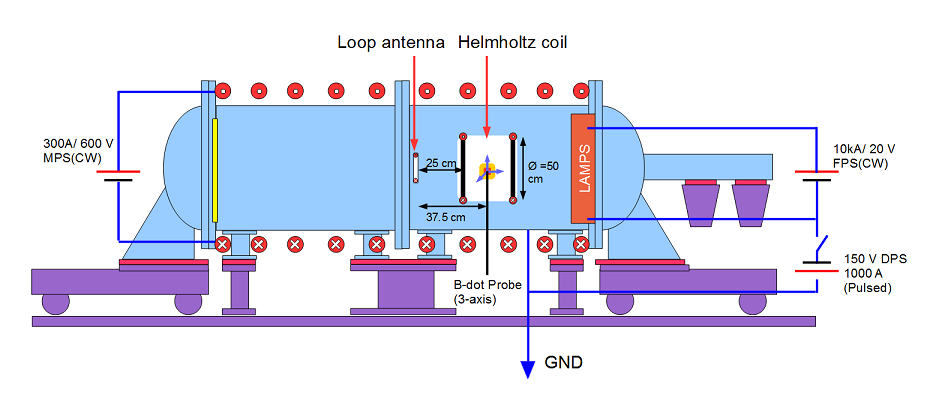}
    \caption{Schematic of the large volume plasma device experimental setup showing the locations for multi-filamentary plasma source along with its electrical circuit for plasma discharge, magnet coils, vacuum ports, end plates consisting of permanent magnets, Helmholtz coil assembly, loop exciter and vacuum ports for diagnostics.}
    \label{fig:lvpd-setup}
\end{figure*}

The large volume plasma device~\cite{1} is a double-walled cylindrical vacuum (plasma) system with dimensions of $3\,\mathrm{m}$ in length and $2\,\mathrm{m}$ in diameter, and has recently been upgraded to LVPD-U~\cite{54} (Figure~1). The system is equipped with $94$ vacuum ports distributed in the axial and azimuthal locations for diagnostic implementation. The system is equipped with $10$ large copper solenoid magnet coils (ETP) to produce a uniform magnetic field of $\sim 6.2\,\mathrm{G}$. The large coils are supported with fiber-reinforced plastic (FRP) structures tied with stainless-steel tie rods. The magnet coils can be operated to produce a DC magnetic field of $200\,\mathrm{G}$; in pulsed mode, a magnetic field of $1\,\mathrm{kG}\sim 0.1\,\mathrm{T}$ can be produced. The magnet coils are powered by a $300\,\mathrm{V}$, $600\,\mathrm{A}$ SCR-controlled 12-pulse power supply that can be operated in constant-current / constant-voltage (CC/CV) mode. The system is equipped with a large-area ($2.54\,\mathrm{m}^2$) multi-filamentary plasma source~\cite{2} (LAMPS), featuring 162 hairpin-shaped tungsten filaments distributed over a circular area with a diameter of $1.8\,\mathrm{m}$ to produce primary energetic electrons for plasma generation. The multi-filament plasma source is powered by a computer-controlled filament power supply with a capacity of $20\,\mathrm{V}$, $10\,\mathrm{kA}$. Plasma discharge is controlled with a pulsed discharge power supply with a capacity of $1\,\mathrm{kA}$, $150\,\mathrm{V}$ for a pulse duration of $\Delta t\sim 50\,\mathrm{ms}$. The system also has 12 automated probe-positioning systems~\cite{10,22} comprising computer-controlled linear probe drives. The experiment is controlled with a timer-based trigger generator circuit, and the signals are processed with separate specialized electronic modules to obtain the ion-saturation current, floating potential, electron temperature, and their fluctuation signals. The signals are acquired with a 40-channel PXIe-based fast data-acquisition system (32 channels of $60\,\mathrm{MS/s}$ and 4 channels of $1.25\,\mathrm{GS/s/ch}$).\vspace{0.25cm}

The device attains an ultimate vacuum of $2.0 \times 10^{-6}$ mbar using a combination of pumping system consisting of three liquid nitrogen ($LN_2$) cooled diffusion pumps ($ 1700 l/min$) with each pump backed by individual rotary pumps of capacity $40 m^3/Hr$, one roots pump ($2590 m^3/Hr$) supported by a rotary pump having a pumping speed of $175 m^3/Hr$. \vspace{0.25cm}

In LVPD-U, plasma is generated by electron impact ionization of neutral Argon $(Ar)$ gas maintained at a fill pressure, P\textsubscript{Ar} of \(4 \times 10^{-4}\) mbar. The primary energetic electrons are emitted from the heated filaments of the  multi-filamentary plasma source. These electrons are accelerated by applying a discharge potential ($V_d\sim -70V$) between the hot filaments (cathode) and the device as the anode. The radial confinement of the plasma is maintained by the uniform axial magnetic field ($Bz \sim 6.2 G$). The plasma is pulsed (\(\Delta t_{\text{plasma}} \approx 10 ms - 55 ms \)) and achieves a density \(n_e \sim 1-2 \times 10^{17} \, \text{m}^{-3}\) with a typical electron temperature \(T_e\) in the range of $2 eV$ to $5 eV$. The typical fluctuations in plasma density ($\delta n_e/n_e$) and potential ($\delta \phi/T_e$) are measured to be  <1\% and 1\% \vspace{0.25cm}

Diagnostics include Langmuir probes for measuring ion saturation current ($I_{sat}$), floating potential ($V_f$) and its fluctuations, electron temperature ($T_e$) and plasma potential $V_p$. $3-axis$ magnet probes are used for the detection of components of electromagnetic waves in $x, y, z$ directions. 

\subsection{Helmholtz Coil}

A Helmholtz coil made up of a stainless steel frame, having a diameter of $0.45m$ is installed in LVPD from the top vacuum port [Fig. 2]. The Helmholtz coil can be moved vertically upward and can rotate in the $x-z$ plane. Each Helmholtz coil arms are wound with $10$ turns of the insulated conductors in $2$ sets of $5$ turns each and are connected in series so that they can be individually powered. A typical $2D$ calibration plot measured with a Gauss probe in a grid of $1cm$ in the $x-z$ plane is shown in figure 3. The plot shows a high degree of uniformity in magnetic field is being produced by the Helmholtz coil. 

\begin{figure}[h!]
    \centering
    \includegraphics[width=\columnwidth]{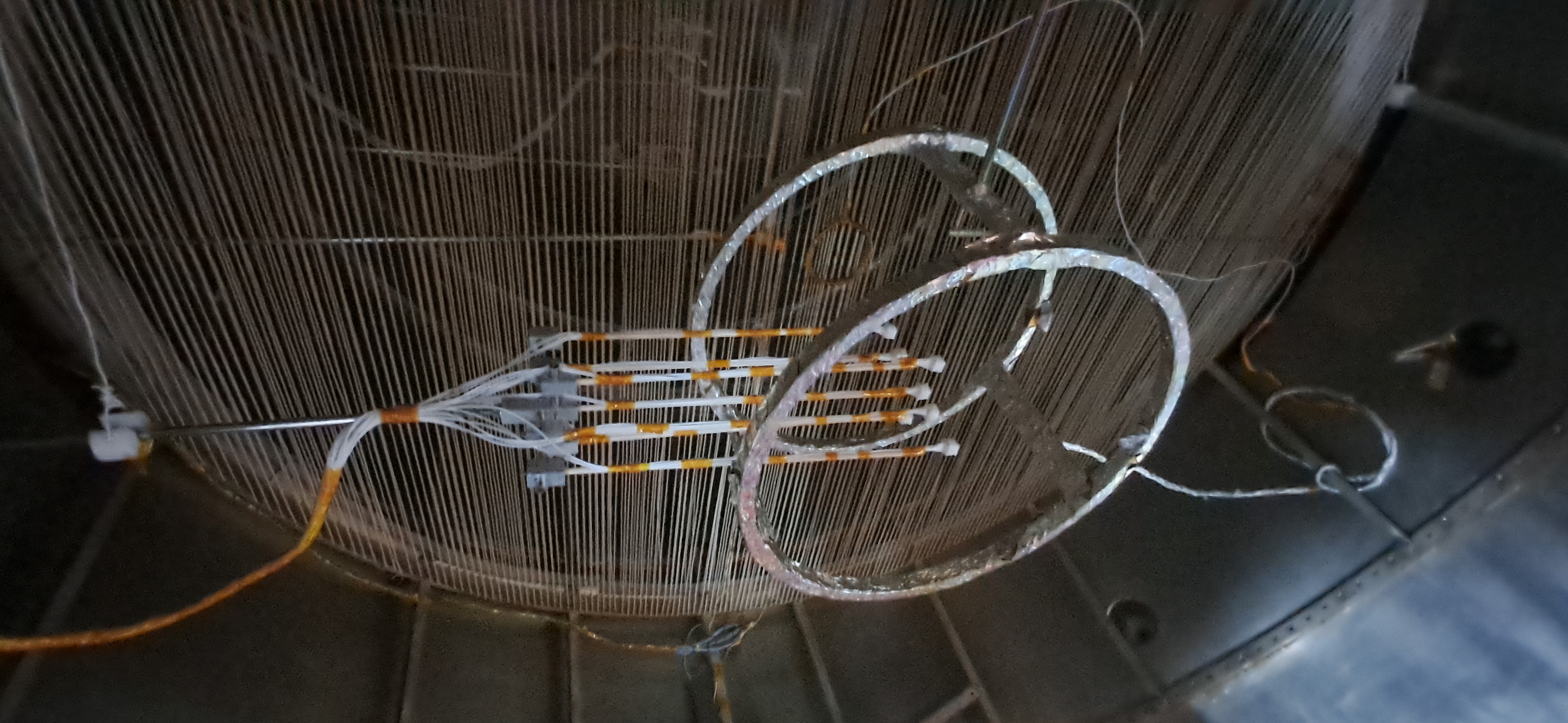}
    \caption{Birds view photograph of Helmholtz coil installed in the axial center of LVPD along with the installed B-dot probe array for the measurement of $k_\parallel$.}
    \label{fig:helmholtz-photo}
\end{figure}
\vspace{0.25cm}

\begin{figure}[h!]
    \centering 
    \hspace*{-0.5cm}
    \includegraphics[width=0.55\textwidth, height=0.35\textheight]{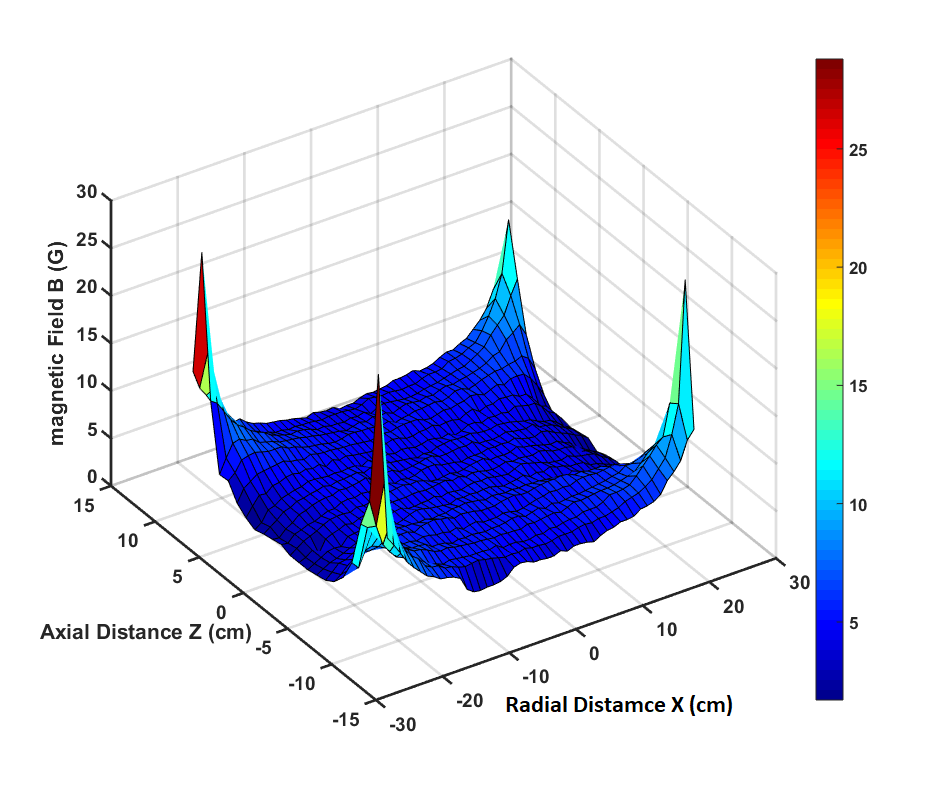} 
    \caption{Surface ($X-Z$) magnetic field profile of Helmholtz coil measured with a Gauss probe.}
    \label{fig:helmholtz-field-xz}
\end{figure}

The magnetic field \( B \) at the center of a Helmholtz coil pair is estimated using the mathematical equation mentioned below. 

\[
B = \frac{8 \mu_0 N I}{\sqrt{125} \, R}
\]

Where,  $\mu_0$  is the permeability in free space (\( \mu_0 = 4\pi \times 10^{-7} \, \text{H/m} \)), \( N \) is the number of coil turns in each Helmholtz coil arm, \( I \) is the current flowing through the coils in series, \( R \) is the radius of the coil. \vspace{0.25cm}



\subsection{Exciter System}
Electromagnetic whistler waves are excited using a loop antenna (L$ \sim 4\mu$H) and an exciter system. The exciter system has incorporated a high voltage pulse power system using a capacitor-based pulse forming network (PFN). This system is operated by discharging a $50 \Omega$ PFN capable of storing up to $12 C$ of charge at $20 kV$ through a triggered thyratron to a loop antenna. The PFN consists of $12$ sections, each comprising a $12.5\mu H$ inductor and a $5 nF$ capacitor. The inductor has a length of $55 cm$ and a diameter of $65 mm$. High-frequency capacitors ($5 nF/30 kV$) are placed along the inductor, spaced at every $12.5 \mu H$ on the solenoid to construct the PFN. \vspace{0.25cm}

A rectangular waveform is generated using the PFN. In addition, to generate a half sinusoidal wave, capacitors of various capacitances have been utilized. Two hydrogen thyratrons (FX2519A, E2V Technologies) work as the switch and the crowbar to achieve the desired pulse width. Reliable and uniform PFN switching, in the range of $500 V$ to $20 kV$, is ensured by controlling the reservoir heater current in the thyratron. The PFN output is connected to the loop antenna via a ferrite core transformer. Impedance matching between the PFN and the load is accomplished by biasing the transformer's core with a DC current. The pulse current to the loop antenna is measured using an Ion-Physics make current transformer (CT) that has a maximum frequency bandwidth of approximately $10 MHz$. Figure 4a shows a typical waveform generated by the PFN. Figure 4b shows the power spectrum of the PFN signal obtained by its Fourier transform, and Figure 4c shows the pulses of the wave magnetic field ($\delta B_x, \delta B_y$ and $\delta B_z$) captured by a 3-axis magnet probe during excitation in plasma.
\begin{figure*}[!t]
    \centering
    \hspace*{-0.40cm}
    \includegraphics[width=0.80\linewidth , height=0.50\textheight]{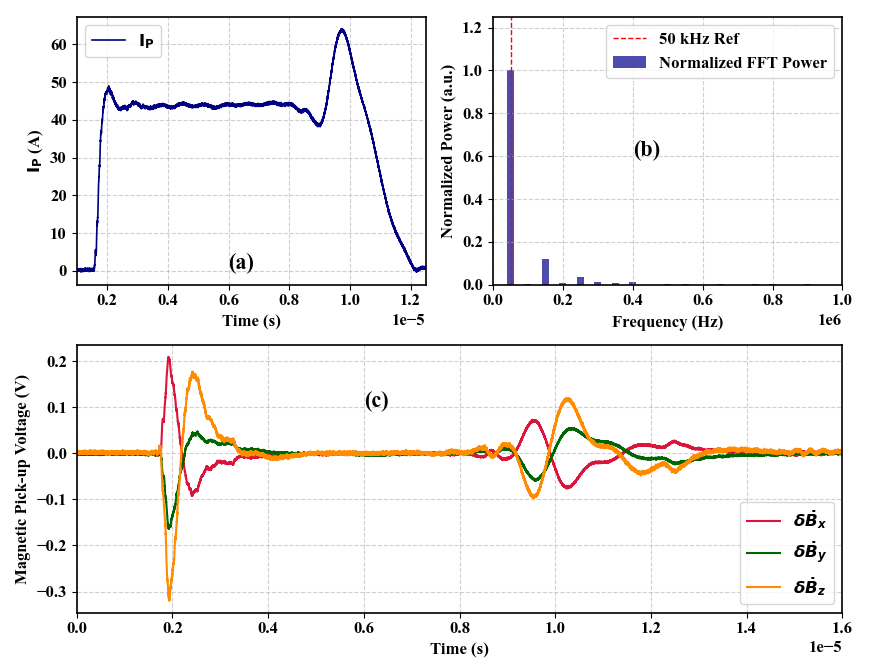}
    \caption{Typical (a) Exciter current pulse measured by CT with discharging PFN, (b) normalized power spectrum of the loop-antenna pulse with respect to maximum power at $50\,\mathrm{kHz}$, and (c) magnetic-signal components ($\delta B_x$, $\delta B_y$, and $\delta B_z$) captured by a 3-axis magnetic probe in plasma.}
    \label{fig:exciter-waveforms}
\end{figure*}





\section{Plasma Characterization}
The experiments in LVPD-U were conducted to characterize the plasma in different magnetic field scenarios produced by combining the magnetic fields of Helmholtz coils and the background magnetic field. Plasma characterization is performed by measuring the radial profiles of the plasma density ($n_e$), electron temperature ($T_e$), floating potential ($V_f$), and plasma potential ($V_p$) radial profiles. The experiments were performed in the presence of (a) an axial magnetic field $B_z=6.2G$, (b) combination of Helmholtz field of $B_H = 20G$ produced along the direction of the axial magnetic field, (c) combination of Helmholtz field of $B_H = -20G$ produced opposite to the direction of the axial magnetic field, and (d) the production of a magnetic NULL region at the center plane of the Helmholtz coil in the presence of plasma using a cylindrical Langmuir probe with a diameter of $1 \, \text{mm}$ and a length of $10 \, \text{mm}$. Measurements were taken radially outward from the center of the coils to $50 \, \text{cm}$ in the $x$ direction. Subsequently, wave-launching experiments were conducted under the experimental conditions mentioned above. 

\begin{figure}[h!]
    \centering
    \hspace*{-0.70cm}
    \includegraphics[width=1.050\linewidth , height=0.33\textheight]{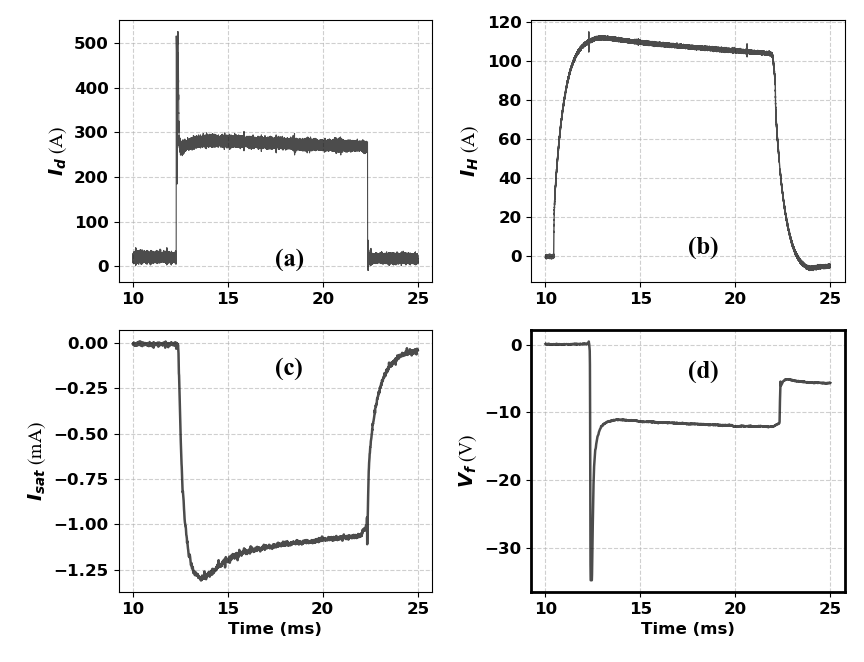}
    \caption{Typical time pulses for (a) plasma discharge current ($I_d$), (b) Helmholtz coil charging current ($I_H$) is shown for charging $20G$, (c) ion saturation current ($I_{sat}$), and (d) floating potential ($V_f$) in plasma. }
    \label{fig: figure }
\end{figure}

 Figure 5a shows a typical pulse for the plasma discharge current ($I_d$) measured during the experiments. The current spike at the onset of plasma discharge is due to the avalanche of primary energetic electrons released from the hot tungsten filaments of multi-filamentary plasma source. Figure 5b shows the charging current ($I_H$) of the Helmholtz coil to produce a magnetic field of $B_H = 20G$. It has a full charging response time of $\sim 2 ms$ due to finite inductance ($\mu_H\sim mH$) of the multi-turn windings in the Helmholtz coil. Figure 5c shows the time evolution of the typical ion saturation current ($I_{sat}$) and fig. 5d shows the typical floating potential profile ($V_f$) measured in the LVPD plasma. During the onset of the plasma, the floating potential is showing a peak value similar to the observations made in the discharge current and is due to the avalanche of primary energetic electrons from the filaments and thereafter reached to a steady state value of floating potential.

\subsection{Plasma Density ($n_e$)}

Figure 6 shows the radial profile of the plasma density ($n_e$) measured using a Langmuir probe starting from the Helmholtz coil center up to a distance of $0.5m$. It should be noted that the Helmholtz coil radius is $0.225m$. Plasma density is estimated by measuring the ion saturation current ($I_{sat}$) and the electron temperature ($T_e$) using Langmuir probes and utilizing the formulation $n_e = \frac{I_{sat}}{0.61eA_pC_s}$. Here, $e$ is the electronic charge, $A_p$ is the probe surface area, and $C_s = \sqrt{\frac{T_e}{m_i}}$ is the acoustic speed of the ions in plasma. It can be seen that, with the increase in the magnetic field value of the Helmholtz coil, the magnitude of the plasma density is increased because of the magnetization of charge particles. However, the plasma density profile is found to be flat in the experimental region in the range of $\sim 0.7\times 10^{17} m^{-3}$  to $\sim 1.2\times 10^{17} m^{-3}$ with a high degree of plasma quiescence level $\frac{\delta n_e}{n_e}\approx 0.1 \%$.

\begin{figure}[h!]
    \centering
    \includegraphics[width=0.51\textwidth, height=0.32\textheight]{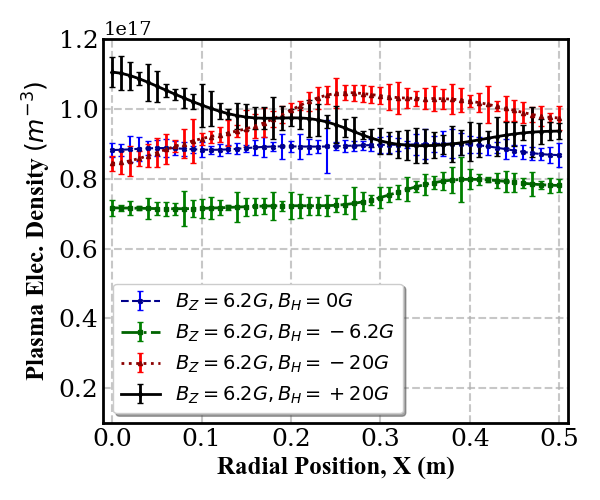}
    \caption{Radial profile for plasma electron density ($n_e$) in different magnetic field conditions produced in the Helmholtz coil setup.}
    \label{fig:enter-label}
\end{figure}

\subsection{Plasma Electron Temperature ($T_e$)}

\begin{figure}[h!]
    \centering
    \includegraphics[width=0.45\textwidth, height=0.32\textheight]{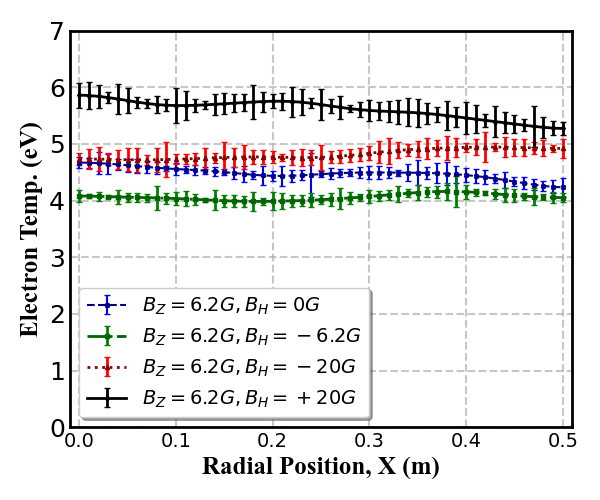}
    \caption{Radial profile for plasma electron temperature ($T_e$) in different magnetic field conditions produced in the Helmholtz coil setup.}
    \label{fig:enter-label}
\end{figure}

Figure 7 shows the radial profile of the plasma electron temperature ($T_e$) measured using the current - voltage characteristics of a Langmuir probe ($I-V$) by biasing the probe in the range of $-70V \le V_B \le 30V$. The electron temperature is estimated from the inverse slope obtained from the logarithmic value of the probe electron current ($I_e$) and the bias voltage ($V_B$). It also can be seen that, as the magnetic field value of the Helmholtz coil is increased, the magnitude of the plasma electron temperature is increased because of the increased confinement of charge particles including hot electrons in the plasma. However, the electron temperature profile is found to be flat in the experimental region in the range of $4eV$ to $6eV$. 

\subsection{Plasma Floating Potential ($V_f$)}

The plasma floating potential ($V_f$) has been measured using the Langmuir probe in floating mode with an impedance of $10M\Omega$. We measured the peak and steady-state floating potential for each measurement (Fig. 5d). Figure 8 shows the radial profiles for plasma floating potential measurements at different magnetic field configurations. The radial floating potential profiles are observed to be flat and in the range of $-12V$ to $-10V$ for all experimental conditions. Here, the peak floating potential showed a variation between the magnetized and non-magnetized plasma conditions within the Helmholtz coil. Outside the Helmholtz boundary the peak floating potential is of similar orders. It can be seen that the peak floating potential value at the Helmholtz coil center for a magnetic null case is high, i.e. less negative, as the energetic electrons are less confined. As the magnetic field value becomes high, the peak of floating potential has started to decrease (increase) towards lower values by virtue of trapping of energetic electrons. However, this effect is not prominent beyond the radial boundaries of the Helmholtz coil (i.e. $22.5cm$). 

\begin{figure}[h!]
    \centering
    \includegraphics[width=0.5\textwidth, height=0.31\textheight]{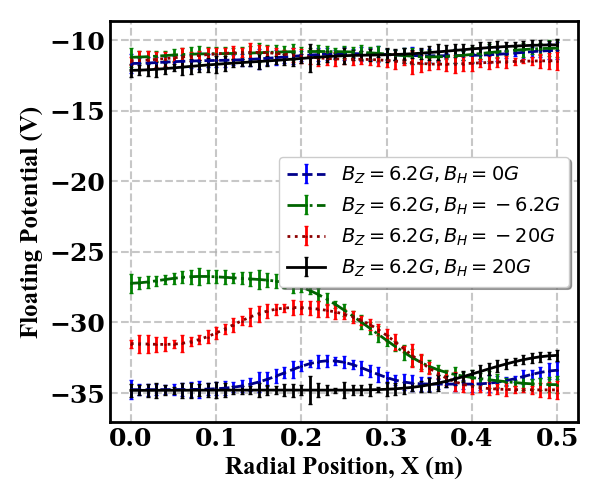}
    \caption{Radial profile for plasma floating potential ($V_f$) in different magnetic field conditions produced in the Helmholtz coil setup.}
    \label{fig:enter-label}
\end{figure}

\subsection{Plasma Potential ($V_p$)}
\begin{figure}[h!]
    \centering
    \includegraphics[width=0.50\textwidth, height=0.31\textheight]{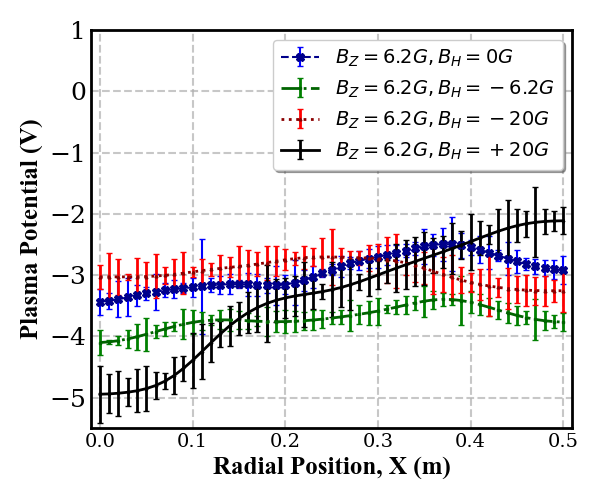}
    \caption{Radial profile for plasma potential ($V_p$) in different magnetic field conditions produced in the Helmholtz coil setup.}
    \label{fig:enter-label}
\end{figure}

Figure 9 shows the radial profile of the plasma potential ($V_p$) measured using the current - voltage characteristics of a Langmuir probe ($I-V$) by biasing the probe in the range of $-70V \le V_B \le 30V$. The plasma potential is estimated from the inflection point technique from the intersection of the slopes of the electron temperature ($T_e$) and electron saturation current ($I_{e,sat}$) region. It can be seen that, for the magnetic null configuration, the radial plasma potential profile is almost uniform. However, in other cases, the plasma potential is also fairly uniform within the Helmholtz coil boundary ($E\approx 0$) except for high magnetic field conditions ($E = 14 V/m$). As the magnetic field value increases the plasma potential becomes more negative probably because of more energetic electron trapping in the magnetic fields and enhanced gyro frequency. The range of plasma potential in the experimental region is found to be in the range of $-5V$ to $-3V$ for different magnetic field configurations. 


The above experimental observations suggest that the plasma is fairly uniform in the experimental region. In the absence of finite gradients in electron density ($\nabla n_e$) and temperature ($\nabla T_e$) from the plasma center to the Helmholtz coil boundaries, free energy sources are not available for coupling during plasma-wave interaction in wave excitation studies. Furthermore, the absence of gradients in plasma pressure profiles ($\nabla P_e = T_e\nabla n_e + n_e\nabla T_e$) does not generate any diamagnetic current in the experimental system; thus, the background magnetic field expulsion in the plasma can be neglected.

\section{Electromagnetic Wave (Whistler) Excitation Experiment}
In LVPD-U, high frequency whistler mode waves are excited by a single turn (Dia: $10cm$) loop antenna made from a teflon coated co-axial silverized copper cable in vacuum and plasma conditions. The loop antenna is excited with pulse current from a high voltage PFN. We conducted several wave excitation and detection campaigns under various charging conditions of Helmholtz coils. In each experimental configuration, we obtained different types of observational data on the $3-$axis magnetic probe. In addition, we performed a fast Fourier transform (FFT) operation on the data set to find the dominant mode frequency excited from the loop antenna and waves propagated in plasma media. We also performed a radial scan of the B-dot probe starting from the center of the Helmholtz coil to the outside covering a radial span of 50 cm towards the LVPD device wall for wave detection. 
First, we performed calibration measurements under vacuum conditions, i.e., in the absence of both plasma and background magnetic field ($B_z=0\mathrm{G}$), to determine the propagation path of the excited electromagnetic (EM) wave in the LVPD-U chamber. Thereafter, measurements were made in plasma in the presence of a uniform axial magnetic field ($B_z=6.2\mathrm{G}$, $B_H=0\mathrm{G}$). In this configuration, two exciter current regimes were investigated: a low exciter current ($I_p=6\mathrm{A}$ and $10\mathrm{A}$, corresponding to $B_{\mathrm{ext}}/B<1$) to study the plasma response to a relatively linear perturbation and a high exciter current ($I_p=50\mathrm{A}$, corresponding to $B_{\mathrm{ext}}/B\geq1$) to investigate the response to a nonlinear perturbation in the presence of only the axial magnetic field.

Finally, at high exciter current ($I_p=50\mathrm{A}$, $B_{\mathrm{ext}}/B\geq1$), measurements were performed in the presence of both the axial magnetic field and the Helmholtz field to investigate the plasma response to a nonlinear perturbation under different magnetic field configurations.

For all of the above configurations, the propagation and electromagnetic-field characteristics of the excited wave, as well as its interaction with the plasma, were measured. The corresponding observations are summarized in this section.

\subsection{Loop Antenna Excitation in Vacuum with $B_z = 0$}
\begin{figure}[h]
    \centering
    \includegraphics[width=1.0\linewidth , height=0.25\textheight]{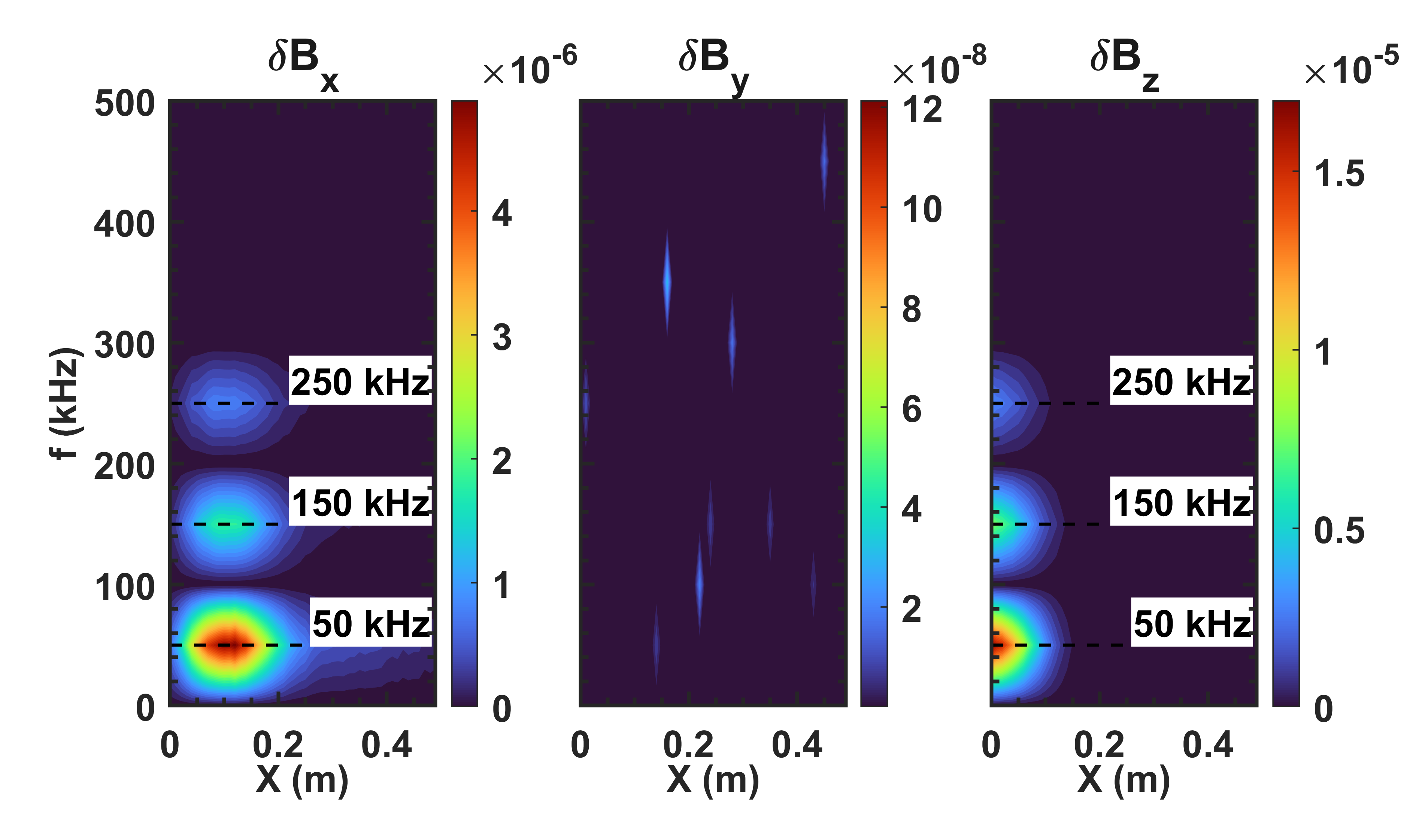}
    \caption{Radial profile for auto power spectrum of magnetic perturbation ($\delta B_x, \delta B_y$ and $\delta B_z$) due to loop antenna excitation measured in vacuum conditions without axial magnetic field ($B_z = 0G$) and Helmholtz magnetic field ($B_H = 0G$).}
    \label{fig:enter-label}
\end{figure}

Figure 10 shows the power spectrum of the pulsed EM wave generated from the loop exciter antenna under vacuum conditions only, without any background magnetic field. Magnetic probe measurements have shown that the peak frequency of the electromagnetic wave is $\approx150 kHz$ with side bands present at $50kHz$ and $250kHz$. We could observe that the peak $B_x$ component of the magnetic field has a spread between $x = 5cm - 15cm$, i.e. a finite dispersion in space is observed. The component $B_y$ is approximately two orders less than the measurable field in the $x$ and $z$ directions and therefore can be neglected. 

\begin{figure}[h!]
    \centering
    \includegraphics[width=1.0\linewidth , height=0.3\textheight]{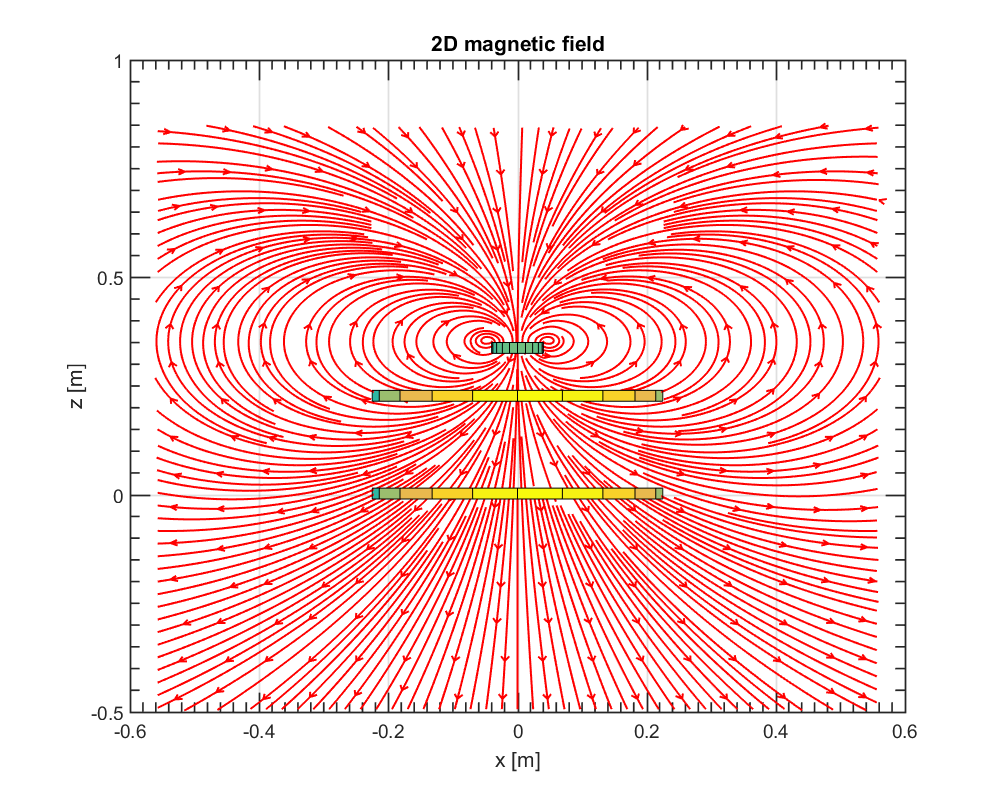}
    \caption{Simulated magnetic field configuration due to the $50A$ excitation of the loop antenna in vacuum conditions without axial magnetic field ($B_z = 0G$) and Helmholtz magnetic field ($B_H = 0G$) (Green Bar = Loop Antenna, Yellow Bar = Helmholtz Coil).}
    \label{fig:enter-label}
\end{figure}

However, the $B_z$ component has shown perfect symmetry in our detection probe with respect to the plasma center. Hence, we conclude that the excited EM wave propagated in the $x-z$ plane as expected from linear theory. The peak $B_x$ component has shown a radial shift probably due to the non-interception of the wave front component at the $z = 0, x = 0$ plane. Figure 11 shows the simulated magnetic field profile (not in scale) in the experimental zone due to the $50A$ excitation of the loop antenna under vacuum conditions without axial magnetic field ($B_z = 0G$), and Helmholtz magnetic field ($B_H = 0G$).

\subsection{Amplitude Scaling: Control Experiment}

A central requirement in identifying harmonic generation as a genuine wave--plasma nonlinearity is to rule out the possibility that the observed spectral features originate from distortions intrinsic to the excitation circuitry itself (e.g.\ thyratron
switching transients, PFN ringing, or saturation in the current transformer). To address this, we performed a dedicated control experiment in which the loop antenna excitation current $I_p$ was varied from its nominal operating value of $50\,\mathrm{A}$ to
$10\,\mathrm{A}$ and $6\,\mathrm{A}$, under identical
plasma conditions ($B_z = 6.2\,\mathrm{G}$, $B_H = 0\,\mathrm{G}$).

Figure~12 shows the measured magnetic pickup waveform (raw and filtered) together with the corresponding
normalized FFT power spectrum for each excitation current. At $I_p = 6\,\mathrm{A}$ and $10\,\mathrm{A}$, the spectrum is dominated almost entirely by the fundamental component near
$50\,\mathrm{kHz}$, with its second-harmonic at $100kHz$. At the nominal excitation current of $I_p = 50\,\mathrm{A}$, the spectrum develops a rich sequence of harmonics extending to several hundred kHz, with comparable power distributed across the $2^{nd}$--$6^{th}$
harmonics.

This dependence on amplitude is consistent with the nonlinearity
parameter of the launched wave, $B_{ext}/B \gtrsim 1$, which is
satisfied only at the highest drive current in our experiment; at
$6$--$10\,\mathrm{A}$ the perturbation remains in the linear,
small-amplitude regime ($B_{ext}/B \ll 1$) and the spectrum is largely dominated by $50kHz$ and $100kHz$, while 150kHz and its higher harmonics are absent.
Since the excitation circuitry
(PFN, thyratron, ferrite transformer, and current transformer) is
operated identically across all three current settings, only the
charging voltage of the capacitor bank is varied, any
harmonics arising from instrumental nonlinearity (e.g.\ core
saturation or switching ripple) would be expected to scale
similarly at all current levels, or at minimum to already be
present at $10\,\mathrm{A}$.

\begin{figure*}[t]
\centering
\includegraphics[width=\textwidth]{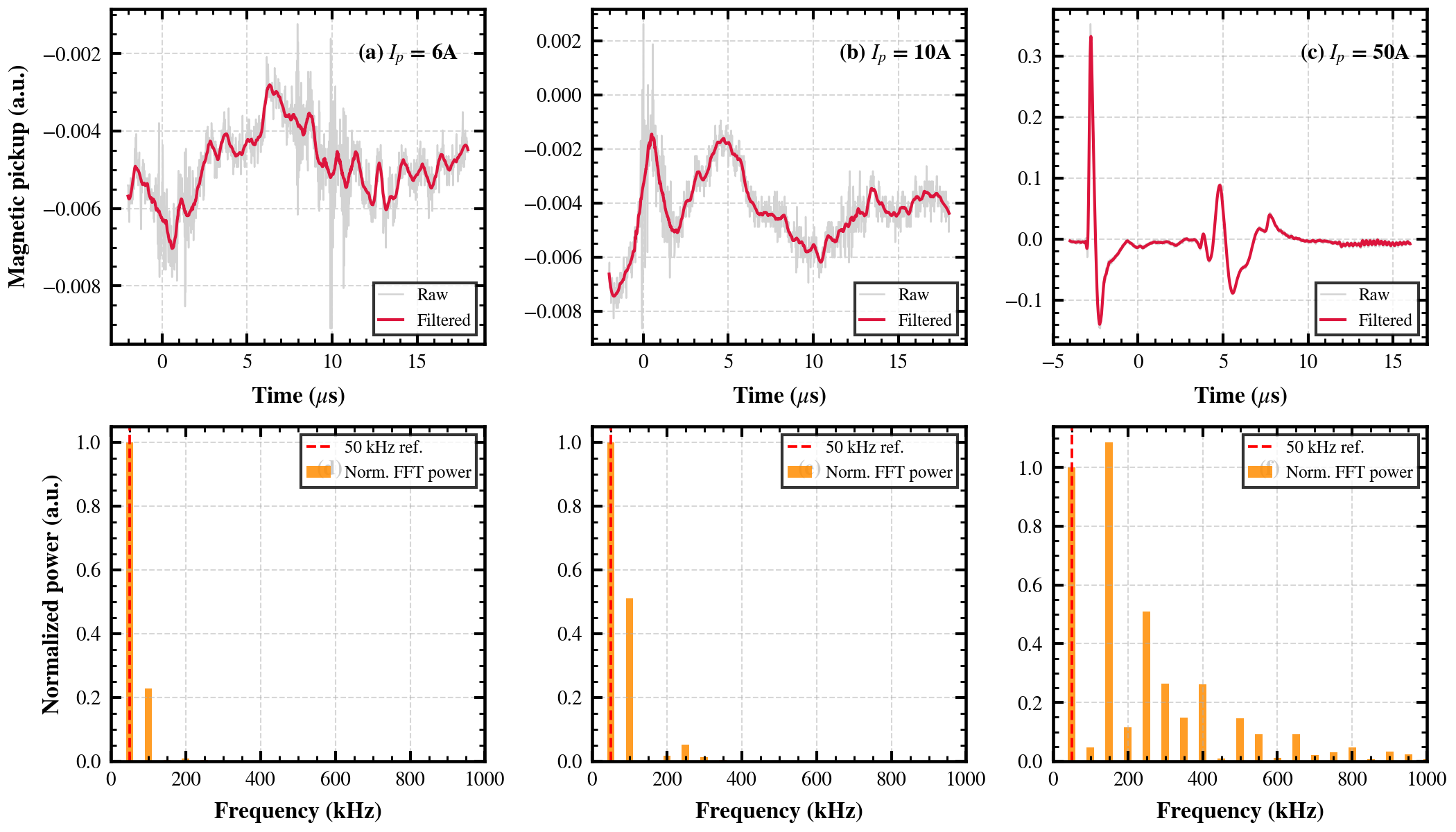}
\caption{
Control experiment demonstrating the dependence of the excited wave spectrum on the excitation current ($I_p$). The upper row shows the measured magnetic pickup signal (gray) together with the filtered waveform (red), while the lower row presents the corresponding normalized FFT power spectra for (a) $I_p=6$ A, (b) $I_p=10$ A, and (c) $I_p=50$ A under identical plasma conditions with $B_z=6.2$ G and $B_H=0$ G.}
\label{fig:current_scan}
\end{figure*}

\subsection{Loop Antenna Excitation in Plasma with $B_z=6.2G,$ $ B_H=20G$}
Figure 13 shows the power spectrum of the pulsed EM wave generated from the loop exciter antenna under plasma conditions with a background magnetic field $B_z = 6.2G$ with Helmholtz field of $B_H = 20G$ applied in the same direction as the background magnetic field. Magnetic probe measurements have shown that the peak frequency of the electromagnetic wave is $\approx 450 kHz$ with $\Delta f/f \approx 1.0 $ in the $B_x$, $B_y$ and $B_z$ components in the presence of plasma. Typical plasma density, $n_e\sim 1.5\times 10^{17} m^{-3}$ and electron temperature, $T_e\sim 3eV - 4eV$ have been maintained during the experiments. The enhancement in frequency by a factor of $\sim 3$ in the detection of peak power has certainly clarified that the wave has interacted with the plasma in the presence of the background magnetic field and the Helmholtz coil field. The simultaneous appearance of $B_x$ and $B_y$ components of the magnetic field suggests polarized EM wave propagation. This observation requires further investigation to understand the underlying physics. The observations also suggest experimental verification of higher-harmonic generation (HHG) in the frequencies in addition to the fundamental frequency of $150kHz$, the appearance of distinct peaks at $300kHz$, $450kHz$, and $600 kHz$, providing clear indication of higher-harmonic generation due to nonlinear wave-plasma interactions, although the spectrum is largely broadband.

\begin{figure}[htbp]
    \centering
    \hspace*{-1.0cm}
    \includegraphics[width=0.58\textwidth, height=0.25 \textheight]{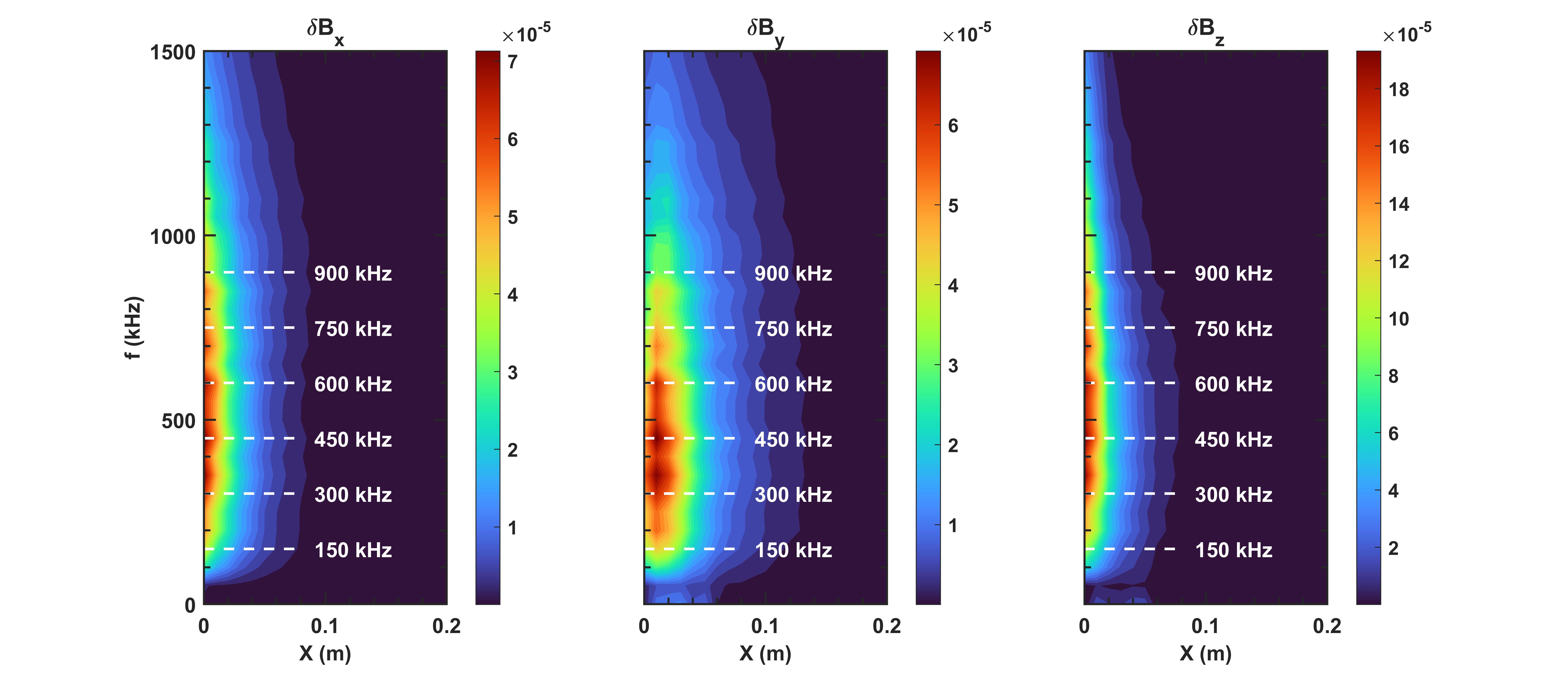} 
    \caption{Radial profile for auto power spectrum of magnetic perturbation ($\delta B_x, \delta B_y$ and $\delta B_z$) due to loop antenna excitation measured in plasma $B_z=6.2G$ and $B_H=20G$.} 
    \label{fig:example}
    \end{figure}

\subsection{Loop Antenna Excitation in Plasma with $B_z=6.2G,$ $B_H=-20G$}

\begin{figure}[h!]
    \centering
    \hspace*{-0.80cm}
    \includegraphics[width=0.58\textwidth, height=0.25 \textheight]{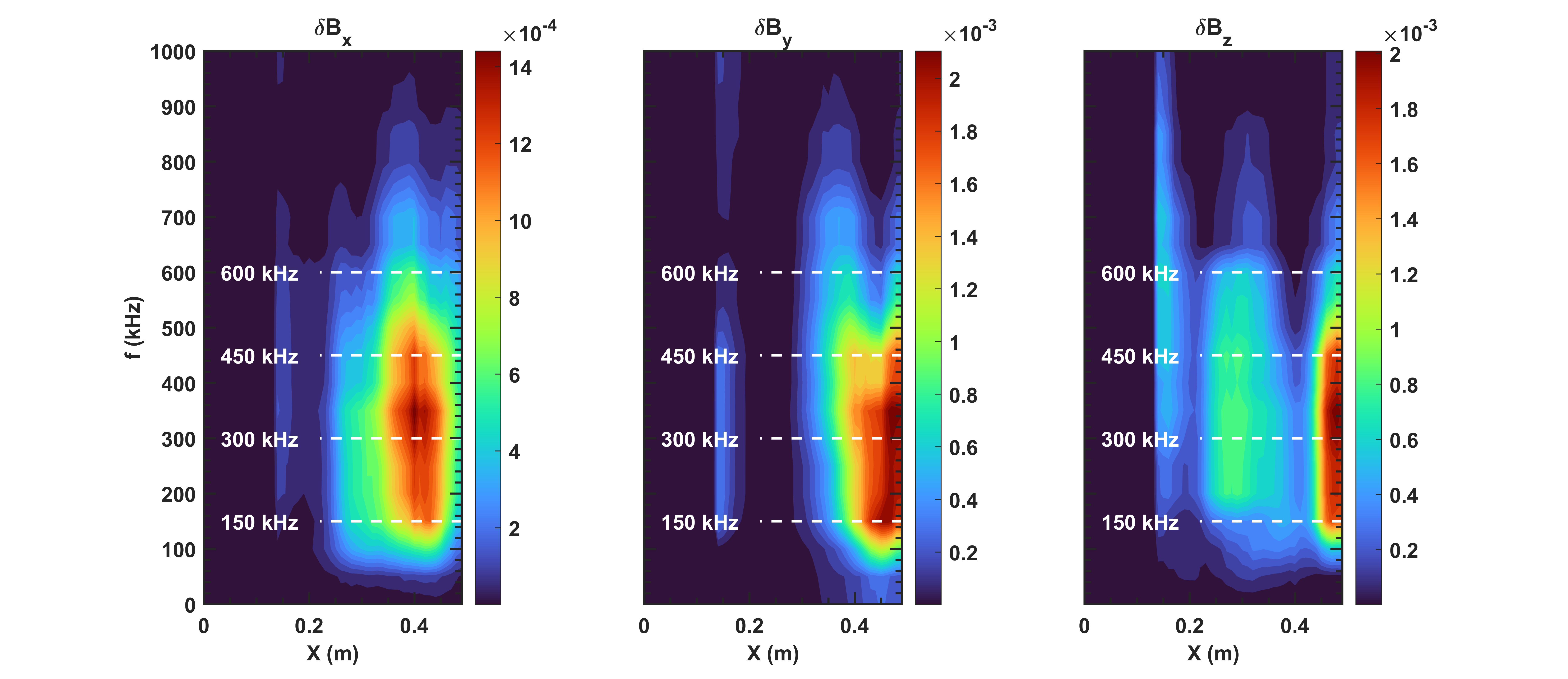} 
    \caption{Radial profile for auto power spectrum of magnetic perturbation ($\delta B_x, \delta B_y$ and $\delta B_z$) due to loop antenna excitation measured in Plasma with $B_z=6.2G$ and $B_H=-20G$.}
    \label{fig:example}
    \end{figure}

Figure 14 shows the power spectrum of the pulsed EM wave generated from the loop exciter antenna under plasma conditions with the background magnetic field $B_z = 6.2G$ and the Helmholtz field of $B_H = -20G$ in the opposite direction to the background magnetic field. The magnet probe measurements have shown the peak frequency of the electromagnetic wave is $\approx 300kHz - 400kHz$ with $\Delta f/f \approx 0.8$ in the $B_x$, $B_y$ and $B_z$ components of the wave magnetic field in the presence of plasma. However, the peculiarity of the observation is that the wave could not be detected within the Helmholtz volume, rather it has propagated outside the Helmholtz system at $x = 40cm - 50cm$. The observed behavior suggests that the present magnetic field configuration may not favor wave propagation within the Helmholtz region. A self-consistent plasma simulation incorporating the magnetic-field configuration is expected to provide further insight into the observed propagation behavior. We also observed higher-harmonic generations for this experimental condition as discussed in the previous subsection. 

\section{Experimental Characterization of the Propagating Whistler Mode}
To establish the identity of the observed electromagnetic fluctuations as whistler waves, a comprehensive experimental characterization is performed. The wave packet is analyzed through its spatio-temporal evolution, propagation delay, dispersion characteristics, phase and group velocities, polarization state, ellipticity, and propagation direction.
The experimentally obtained parameters are subsequently compared with the predictions of the cold-plasma whistler dispersion relation to assess the consistency between theory and measurements.

\subsection{Wave Propagation and Group Velocity}

The arrival time of the coherent wave packet was measured successively at each probe location. The propagation delay between adjacent probes was determined from the temporal shift of the wave packet, allowing the propagation velocity to be estimated. Representative waveforms recorded by the five-probe array are shown in Fig.~15.
\begin{figure}[h!]
    \centering
    \hspace*{-0.45cm}
    \includegraphics[width=0.45\textwidth, height=0.31\textheight]{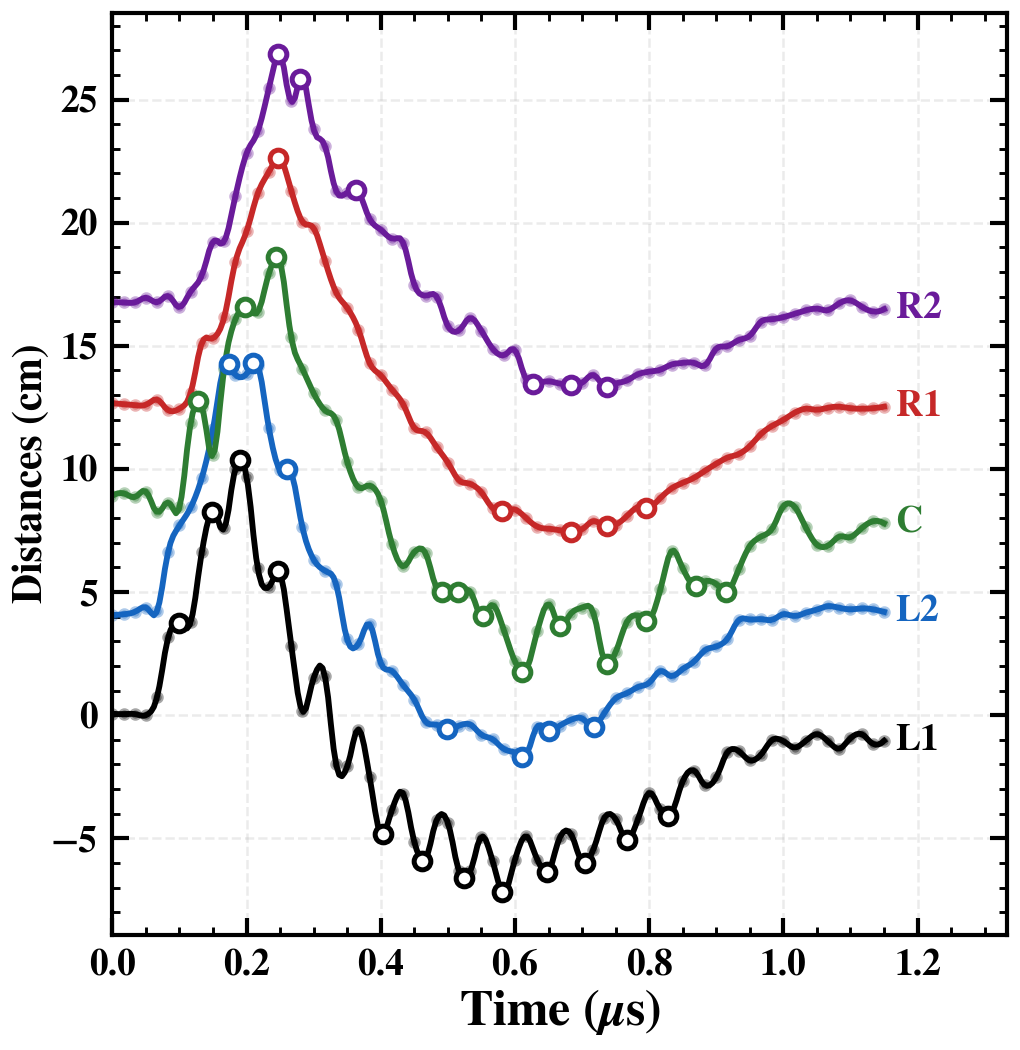}
    \caption{Coherent whistler waveforms measured at different probe locations using a five-element probe array with an inter-probe spacing of 5 cm.}
    \label{fig:enter-label}
\end{figure}

To obtain a more accurate estimate of the propagation delay, cross-correlation functions were calculated between signals recorded by different probe pairs. The measured correlation functions demonstrate strong spatial coherence of the propagating wave throughout the probe array (Fig~16).
\begin{figure}[t]
    \centering
    \includegraphics[width=\columnwidth]{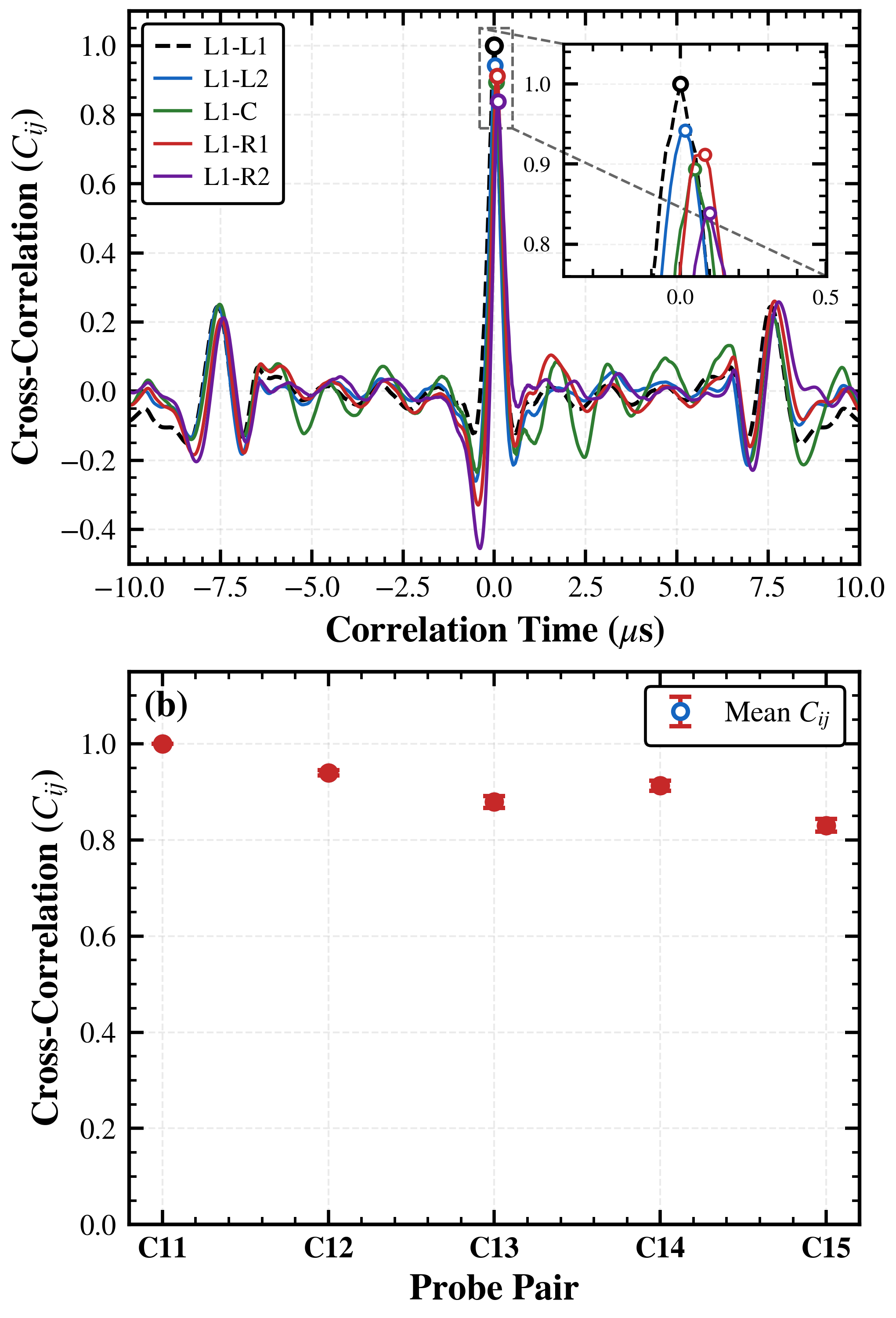}
    \caption{Cross-correlation between magnetic signals measured at spatially separated magnetic probes. (a) Normalized cross-correlation $C_{ij}(\tau)$ as a function of correlation time for probe pairs relative to the reference probe L1; the inset shows a magnified view of the peak region. (b) Ensemble-averaged peak correlation coefficient for each probe pair, with error bars denoting the standard deviation over the shot ensemble.}
    \label{fig:cross_correlation}
\end{figure}

\subsection{Phase velocity and Wave number}
\subsubsection{Experimental determination of the parallel phase velocity and wave number}

The measured magnetic fluctuation signals were transformed into the frequency domain using the FFT. Individual harmonic components were subsequently isolated using narrow-band filtering and transformed back into the time domain. The delay in the propagation of each isolated harmonic between different probes enabled the estimation of the corresponding phase velocity ($V_{ph}$).  The parallel wave number was finally obtained from

\[
k_{\parallel}=\frac{\omega}{V_{ph}},
\]

where $\omega$ is the angular frequency of the selected harmonic. The measured values of $V_{ph}$ and $k_{\parallel}$ were averaged over more than 100 independent plasma shots along with its statistical uncertainties (Fig.~17). The parallel wave numbers experimentally measured at different harmonic frequencies are summarized in Table I.

\begin{figure}[h!]
    \centering
    \includegraphics[width=0.52\textwidth, height=0.30\textheight]{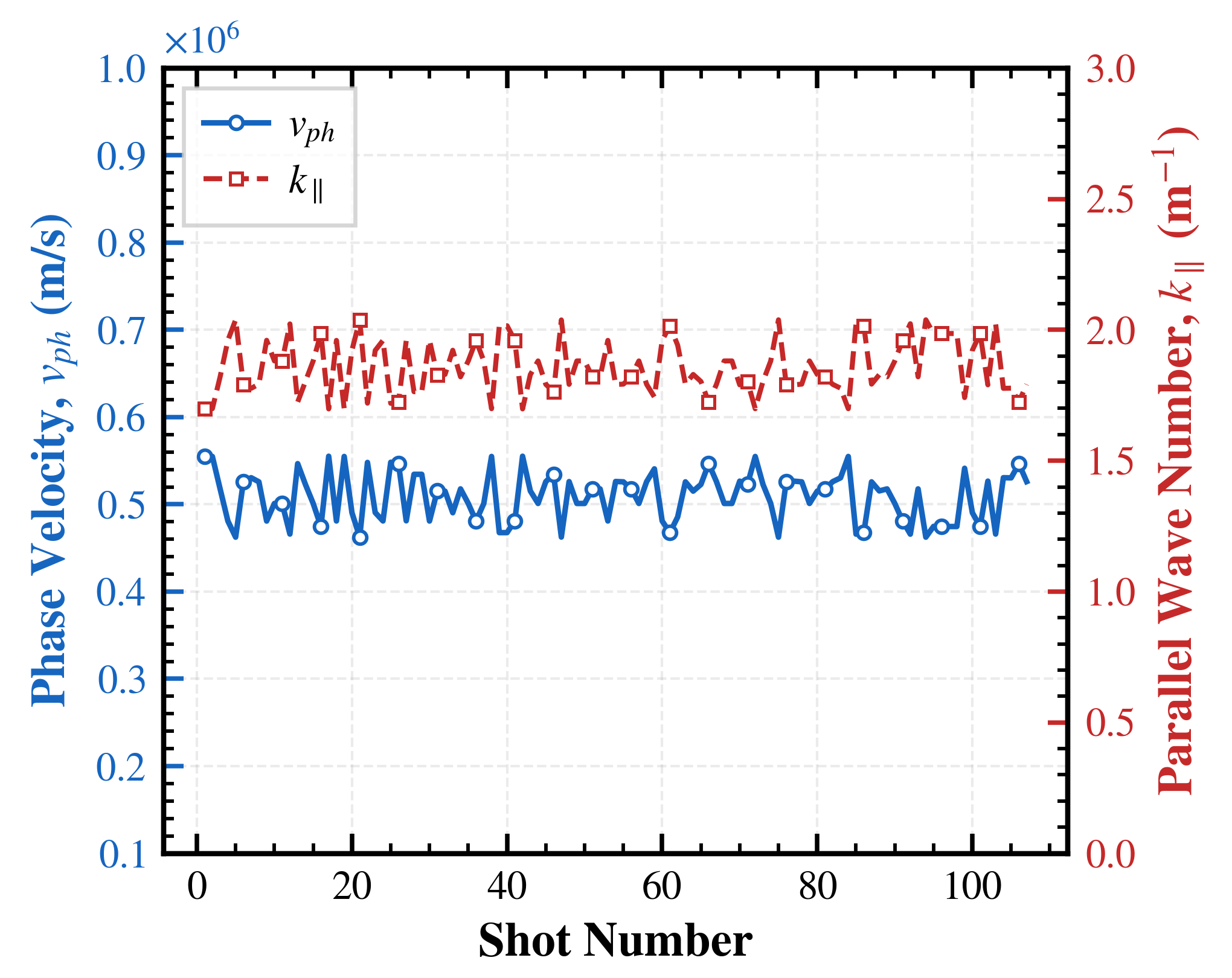}
    \caption{Experimentally measured parallel phase velocity and Parallel wave number at $f_0 =150kHz$ (fundamental frequency) taken over 100+ identical plasma shots.}
    \label{fig:enter-label}
\end{figure}

\subsubsection{Experimental determination of the perpendicular phase velocity and wave number}

To further characterize the propagation properties of the observed waves in the radial direction, the transverse wave number was determined by spatial phase mapping using a single movable probe referenced to the exciter trigger level across highly reproducible plasma discharges, following the approach demonstrated by Santolik(2003)~\cite{59}, Means(1972)~\cite{60} and Stenzel (1993)~\cite{63}. Since all magnetic field measurements were synchronized with the same antenna excitation pulse, the recorded waveforms remained phase coherent across the radial direction.

\begin{equation}
k_{\perp}=\frac{\partial \phi}{\partial x},
\label{eq:kperp}
\end{equation}

where $\phi$ is the measured wave phase and $x$ denotes the position along the radial direction (X).

For the fundamental propagating component at 150~kHz, the measured transverse wave number is $k_{\perp}=2.06~\mathrm{m^{-1}}$, while the independently measured parallel component is $k_{\parallel}=1.87~\mathrm{m^{-1}}$. The normal wave angle with respect to the background magnetic field is therefore estimated as

\begin{equation}
\theta=\tan^{-1}\left(\frac{k_{\perp}}{k_{\parallel}}\right),
\label{eq:theta}
\end{equation}

yielding $\theta \approx 48^{\circ}$ for the 150~kHz wave. Similar values obtained for the higher harmonic components indicate that the observed waves propagate at finite angles with respect to the ambient magnetic field, confirming their oblique propagation characteristics. 

\subsection{Dispersion Analysis}
Experimentally measured total wave numbers were compared with the theoretical cold-plasma oblique whistler dispersion relation calculated using the independently measured plasma parameters ($B_{0}=B_Z +B_H = 26.2 G$ and $n_{e}=1.15\times10^{17}$ m$^{-3}$) at the oblique angle of $\theta=48^{\circ}$ and at $0^{\circ}$ (Fig.18). The measured dispersion follows the theoretical whistler dispersion reasonably well within the experimental uncertainty. The propagation and wave-mode characterization presented in this section corresponds to the plasma condition \(B_z=6.2~{\rm G}\), \(B_H=20~{\rm G}\), giving a local background magnetic field \(B_0=26.2~{\rm G}\) in the measurement region.

For oblique propagation, the cold-plasma whistler dispersion relation is given by

\begin{equation}
\label{eq:oblique-whistler-dispersion}
\frac{c^{2}k^{2}}{\omega^{2}} = \frac{\omega_{pe}^{2}}{\omega\bigl(\omega_{ce}\cos\theta-\omega\bigr)}.
\end{equation}

where $\omega_{ce}=eB_{0}/m_{e}$ is the electron cyclotron frequency,
$\omega_{pe}=\sqrt{n_{e}e^{2}/\varepsilon_{0}m_{e}}$ is the electron plasma frequency, and $\theta$ is the propagation angle.

Experimentally, \(k_\parallel\) could be determined up to 900 kHz from axial propagation measurements, while  the determination of \(k_\perp\) was experimentally limited to frequencies up to 500 kHz. Consequently, the experimentally determined total wavenumber, \(k_{\rm total}=(k_\parallel^2+k_\perp^2)^{1/2}\), is shown (Fig.18) only up to 500 kHz. For comparison, the theoretical Stix dispersion relation is also evaluated using the experimentally measured frequency-dependent propagation angle \(\theta(f)\).

\begin{table}[h]
\centering

\caption{Experimentally measured parallel wave numbers obtained from propagation delay measurements using the linear magnetic-probe array.}
\begin{tabular}{ccc}
\hline
Frequency (kHz) & $k_{\parallel}$ (m$^{-1}$) & $\Delta k$ (m$^{-1}$)\\
\hline
150 & 1.865 & $\pm$0.243\\
250 & 3.180 & $\pm$0.550\\
300 & 3.580 & $\pm$0.243\\
450 & 5.250 & $\pm$0.800\\
500 & 6.030 & $\pm$0.780\\
600 & 6.631 & $\pm$1.280\\
650 & 6.970 & $\pm$0.570\\
750 & 8.330 & $\pm$0.630\\
900 & 9.260 & $\pm$0.170\\
\hline
\end{tabular}
\end{table}

\begin{figure}[t]
    \centering
    \includegraphics[width=\columnwidth]{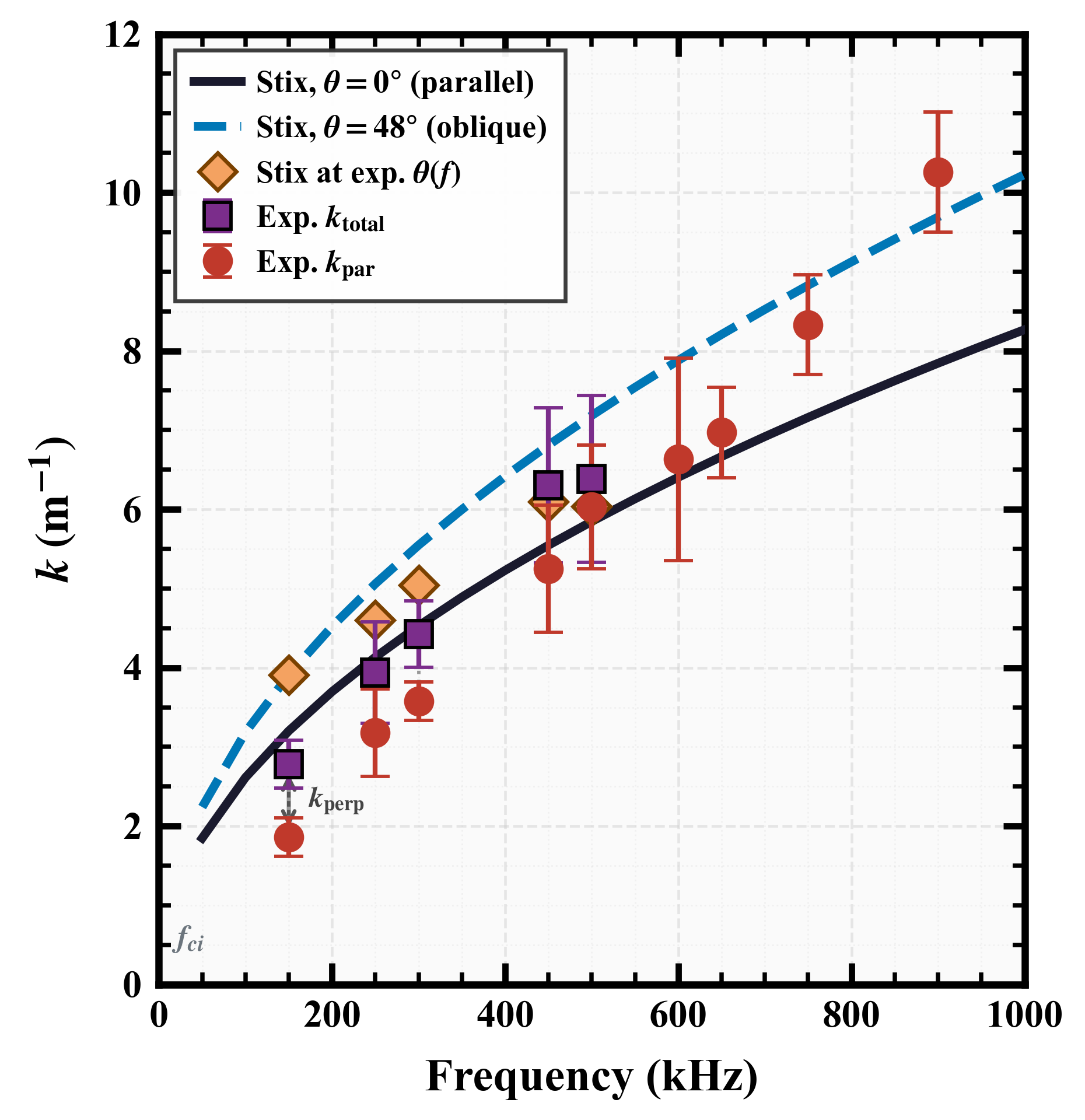}
    \caption{Dispersion relation of the experimentally observed propagating waves. The solid black and dashed blue curves represent the Stix dispersion relation for parallel \((\theta=0^\circ)\) and oblique \((\theta=48^\circ)\) propagation, respectively. The red circles show the experimentally determined parallel wavenumber \(k_\parallel\), measured up to 900 kHz, while the purple squares denote the experimentally determined total wavenumber \(k_{\rm total}=(k_\parallel^2+k_\perp^2)^{1/2}\), limited to frequencies up to 500 kHz. The yellow diamonds represent the Stix dispersion evaluated using the experimentally measured frequency-dependent propagation angle, \(\theta_{\rm exp}(f)\).}
    \label{fig:dispersion}
\end{figure}
\subsection{Polarization Characteristics of the Electromagnetic Fluctuations}

Figure 19 presents the frequency dependence of the polarization ellipticity measured at different radial locations. The polarization ellipticity $\varepsilon$ was calculated from the two transverse magnetic-field components, $\delta B_x$ and $\delta B_y$, as
\begin{equation}
\varepsilon(f) = \frac{|\delta B_x(f)|}
{|\delta B_y(f)|},
\label{eq:ellipticity}
\end{equation}
where $|\delta B_x|$ and $|\delta B_y|$ represent the spectral amplitudes of the corresponding magnetic-field components at the frequency of interest. Thus, $\varepsilon=0$ corresponds to equal amplitudes of the two transverse components, whereas $|\varepsilon|\rightarrow1$ indicates a strongly linearly polarized response dominated by one component. The sign of $\varepsilon$ specifies the relative dominance of the two components according to the adopted coordinate convention.

The measured ellipticity remains positive and finite throughout the investigated frequency range, indicating that the observed waves retain an elliptically polarized character. Although the ellipticity exhibits a radial dependence, no reversal of polarization handedness is observed, and the measured values are consistent with the expected polarization characteristics of obliquely propagating whistler-mode waves.

\begin{figure}[h!]
\centering
\includegraphics[width=0.45\textwidth]{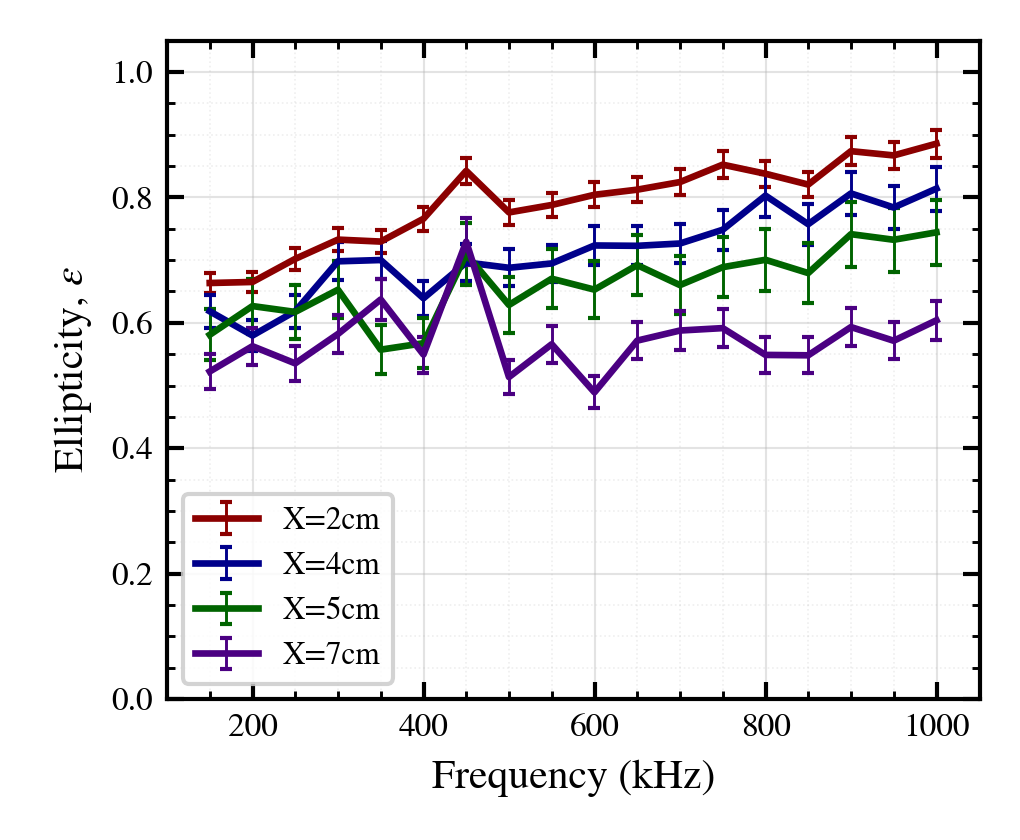}
\caption{Ellipticity spectrum $\varepsilon$ as a function of frequency, measured at $X = 0$--$7$,cm along the LVPD axis. Error bars reflect a $7$--$11\%$ uncertainty in $\varepsilon$ per shot (see text).}
\label{fig:ellipticity}
\end{figure}

\subsection{Bi-coherence analysis of observed modes}
The squared bicoherence spectrum $b^2(f_1,f_2)$ calculated with the method given by Kim (1985)~\cite{64} reveals several localized regions of enhanced nonlinear three-wave coupling rather than a broadband or random distribution, indicating that the observed spectral content is generated through phase-coherent triadic interactions among the two pump wave frequencies (50 and 150~kHz) and their respective harmonics. The squared bicoherence is defined as

\begin{equation}
b^2(f_1,f_2)=
\frac{
\left|
\left\langle
X(f_1)X(f_2)X^*(f_1+f_2)
\right\rangle
\right|^2
}{
\left\langle |X(f_1)X(f_2)|^2\right\rangle
\left\langle |X(f_1+f_2)|^2\right\rangle
},
\label{eq}
\end{equation}
where $X(f)$ represents the complex Fourier transform of the fluctuating signal at frequency $f$, the asterisk ($*$) denotes the complex conjugate, and $\langle \dots \rangle$ indicates an ensemble average over multiple independent data realizations. The numerator corresponds to the squared magnitude of the auto-bispectrum $B(f_1,f_2) = \langle X(f_1)X(f_2)X^*(f_1+f_2) \rangle$, which measures the degree of phase coherence among the triad satisfying the frequency-matching condition $f_3 = f_1 + f_2$. The denominator provides a statistical normalization such that $0 \le b^2(f_1,f_2) \le 1$, where $b^2 \to 1$ indicates complete phase coupling (a coherent nonlinear interaction) and $b^2 \to 0$ signifies independent or randomly phased modes.

The strongest coupling ($b^2 > 0.7$) appears at low $f_1$ across a broad range of $f_2$. Additional coherence islands near $f_1 \approx 250$--$400$~kHz paired with $f_2 \approx 700$--$800$~kHz are consistent with sum-frequency coupling among the pump wave combination of, e.g., $2f_{01} + 3f_{02}$-type. Together, these features support a picture in which the 
higher-harmonic and combination-frequency content in the LVPD spectrum arises from three-wave mixing of the driven pump waves. Broadly, this coupling could be expressed as
\begin{equation}
f_{\alpha\beta}=(\alpha\times50\pm\beta\times150)\,\mathrm{kHz},
\end{equation}
where $\alpha=1,2,3,..$ and $\beta=1,2,3,..\ldots$. 
\begin{figure}[t]
    \centering
    \includegraphics[width=\columnwidth]{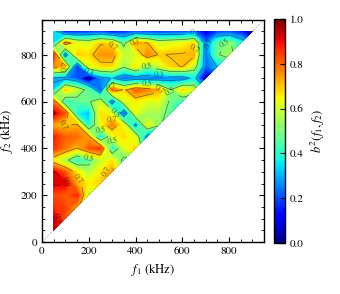}
    \caption{Squared bicoherence spectrum $b^2(f_1,f_2)$ computed from the magnetic 
    fluctuation data, quantifying the degree of phase coupling between spectral components at $f_1$ and $f_2$. Regions of $b^2 \gtrsim 0.5$--$0.8$ (red--orange) indicate statistically significant three-wave nonlinear coupling, with the strongest interactions occurring at low $f_1$ paired with a broad range of $f_2$ and at sum-frequency combinations of the driven pump modes (50, 150, and 250~kHz). These features provide evidence for phase-coherent nonlinear mode coupling among the observed spectral components and are consistent with three-wave interactions.}
    \label{fig:bicoherence}
\end{figure}

\section{Mechanisms of Higher Harmonics Generation}
As discussed in Section $IV$, we have detected higher harmonics of excited fundamental waves in their integral and non-integral forms. This type of nonlinear shift in frequency may be considered as a direct consequence of nonlinear wave-particle and wave-wave interactions in a plasma. The clear physical mechanisms of higher-harmonic generation for electromagnetic waves in plasmas are still an open topic of research, but according to the available liter0ature, finite density gradients and magnetic field anomalies are responsible for generating free energy sources, thus leading to HHG and observation of frequency shift. Apart from these, there are several theoretical frameworks proposed and mathematical tools present to understand the nonlinear interactions in plasma. One such framework is the dielectric approach; in this approach we explore the phenomenon of using dielectric properties of plasma. 
\subsection{Dielectric Approach}
One of the effective tools to understand nonlinear plasma physics is multiscale analysis, in which we expand a physical quantity in terms of a perturbation parameter strength, $\zeta$.

\begin{equation}
    \bm{E}(\bm{r}, t) = \bm{E}_1(\bm{r}, t) + \zeta \bm{E}_2(\bm{r}, t) + \mathcal{O}(\zeta^2),
\end{equation}
\begin{equation}
    \bm{B}(\bm{r}, t) = B_0 \hat{z} + \zeta \bm{B}_1(\bm{r}, t) + \zeta^2 \bm{B}_2(\bm{r}, t) + \mathcal{O}(\zeta^3),
\end{equation}
Here, the background magnetic field is along the $z$-axis, ($B_0 \hat{z}$).
Additionally, $\bm{E}_1(\bm{r}, t)$ and $\bm{B}_1(\bm{r}, t)$ correspond to fundamental harmonic perturbations, while $\bm{E}_2(\bm{r}, t)$ and $\bm{B}_2(\bm{r}, t)$ correspond to the second harmonic.

Putting Equations $(5)$ and $(6)$ in the Maxwell equations~\cite{35} and equating coefficients of all powers of $ \zeta $ we shall get the fundamental and higher harmonics. Here, we limit our mathematical analysis to the second harmonics; the description of other higher harmonics, and detailed numerical analysis of all other higher harmonics will be the subject of other papers.
\begin{align}
\nabla \times \nabla \times \bm{E}_1(\bm{r}, t) 
&+ \frac{1}{c^2} \frac{\partial^2 \bm{E}_1(\bm{r}, t)}{\partial t^2} \nonumber \\
&+ \frac{4\pi}{c^2} \frac{\partial}{\partial t}
\left[ \bm{\sigma}_1(\bm{r}, t) * \bm{E}_1(\bm{r}, t) \right] = 0
\end{align}

\begin{align}
\nabla \times \nabla \times \bm{E}_2(\bm{r}, t) 
&+ \frac{1}{c^2} \frac{\partial^2 \bm{E}_2(\bm{r}, t)}{\partial t^2} \nonumber \\
&+ \frac{4\pi}{c^2} \frac{\partial}{\partial t}
\left[ \bm{\sigma}_1(\bm{r}, t) * \bm{E}_2(\bm{r}, t) \right] \nonumber \\
&= -\frac{4\pi}{c^2} \frac{\partial}{\partial t}
\left[ \bm{\sigma}_2(\bm{r}, t) * \bm{E}_1(\bm{r}, t) \right]
\end{align}

In Fourier form, the equations become the following:

\begin{equation}
\left[
\bm{\epsilon}(\omega_1,\bm{k}_1)
- \frac{c^2 k_1^2}{\omega_1^2} \bm{I}
+ \frac{c^2 \bm{k}_1 \bm{k}_1}{\omega_1^2}
\right]
\cdot \bm{E}(\omega_1,\bm{k}_1) = 0
\end{equation}

\begin{equation}
\left[
\bm{\epsilon}(\omega_2,\bm{k}_2)
- \frac{c^2 k_2^2}{\omega_2^2} \bm{I}
+ \frac{c^2 \bm{k}_2 \bm{k}_2}{\omega_2^2}
\right]
\cdot \bm{E}(\omega_2,\bm{k}_2)
= -\frac{4\pi i}{c^2} \bm{J}_{\mathrm{NL}}
\end{equation}
Here, $\epsilon_1(\bm{k}, \omega)$ and $\epsilon_2(\bm{k}, \omega)$ are the linear and nonlinear dielectric tensors, and $\bm{J}_{NL}(\bm{k}, \omega)$ is the nonlinear current density. The nonlinear current $\bm{J}_{NL}(\bm{k}, \omega)$ can drive the second harmonic, even if it falls outside the inherent wave modes of the plasma system.

Additionally, the expression nonlinear current density is given by:
\begin{equation}
    \bm{J}_{NL}(\bm{k}, \omega) = \sum_j \int q_j \bm{v} F_2(\bm{r}, t, \bm{v}) d^3v,
\end{equation}

Here, the summation is done over all particle species.

\vspace{0.15cm}

The source of second harmonics is considered to be the nonlinear current density. This is understood as, when the first-order density $( \bm{n}_{1}(\bm{r}, \bm{t}))$ and velocity $( \bm{v}_{1}(\bm{r}, \bm{t}))$ perturbations are such that their product is finite, then it will further contribute to induce the nonlinear current density $(\bm{J}_{NL}(\bm{k}, \omega))$ in plasmas, and that becomes the source of second harmonics.
Figure 20 shows the normalized power of the higher harmonics generated corresponding to the pump wave frequency $f_{0} = 150kHz$ in plasmas, extending up to the $8^{th}$ harmonic. A numerical verification of these harmonics will be carried out by solving the dispersion relation discussed above, which will be addressed in a future work.
\begin{figure}[h!]
    \centering
    \hspace*{-0.50cm}
    \includegraphics[width=1.1\linewidth , height=0.35\textheight]{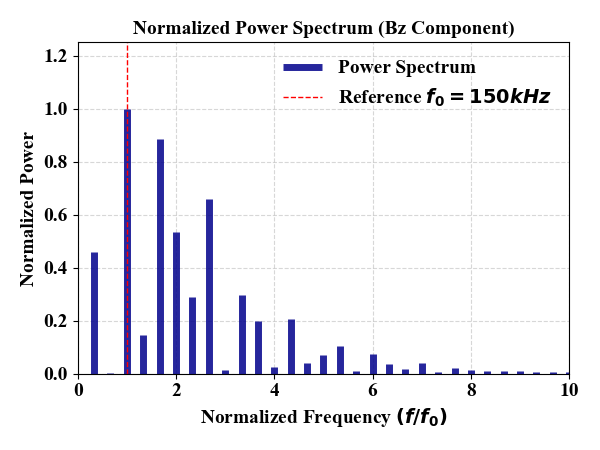}
    \caption{Normalized power spectrum of HHG with corresponding fundamental frequency at $f_0 = 150 kHz $ for Shot no. $\# 51818$, $B_H = 20G, B_z = 6.2G$.}
    \label{fig:enter-label}
\end{figure}

\subsection{Harmonic Generation - Spectral Power Conversion Efficiency}
In this section, we emphasize the spectral power conversion efficiency of various types of higher harmonics ($2^{nd},3^{rd}$, etc.) generated with respect to the fundamental frequency at $150 kHz$. The conversion efficiency of $n^{th}$ harmonics with respect to the fundamental harmonics is given as;
\begin{equation}
    \eta_{n} = \lvert \frac{B(n\omega_0)}{B(\omega_0)}\rvert ^2
\end{equation}

Using Eq. 12, we estimate the conversion efficiency of the spectral power density corresponding to the higher harmonics generated, where $|B(n\omega_0)|^2$ and $|B(\omega_0)|^2$ are the power spectra of the $n^{th}$ harmonics and the fundamental harmonics, respectively, in the frequency space. The noteworthy aspect of these observations is the simultaneous occurrence of harmonics in all magnetic field components, i.e., in $\delta B_x, \delta B_y,$ and $\delta B_z$. The transverse component, $\delta B_{y}$, which emerges in the presence of plasma, also exhibits higher-order harmonic generation with appreciable conversion efficiency. Table II presents the spectral power conversion efficiency of harmonics up to the sixth order for all three magnetic field components, as observed in our measurements. It is particularly significant that the $z-$ component shows a spectral power conversion efficiency comparatively higher than the other components.

\begin{table}[htbp]
\caption{Estimated spectral power conversion efficiency for higher-harmonic generation.}
\centering
\small
\setlength{\tabcolsep}{10pt}      
\renewcommand{\arraystretch}{2.5} 
\begin{tabular}{c c c c c c c}
\hline\hline
\multicolumn{2}{c}{\textbf{Harmonics}} 
& \textbf{$2^{nd}$} 
& \textbf{$3^{rd}$} 
& \textbf{$4^{th}$} 
& \textbf{$5^{th}$} 
& \textbf{$6^{th}$} \\ 
\hline

\multirow{3}{*}{\rotatebox{90}{\textbf{Efficiency (\%)}}}
& $\eta_{x}$ & 38.5 & 0.60 & 1.44 & 4.65 & 4.93 \\

& $\eta_{y}$ & 29.39 & 0.44 & 0.80 & 2.38 & 2.31 \\

& $\eta_{z}$ & 53.27 & 1.30 & 2.71 & 7.04 & 7.56 \\

\hline\hline
\end{tabular}
\end{table}

\section{Conclusion}
In this work, we experimentally investigated the excitation, propagation, and nonlinear evolution of electromagnetic waves in the whistler-frequency regime in the Large Volume Plasma Device-Upgrade (LVPD-U). A clear distinction was observed between wave excitation in vacuum and wave propagation in plasma. Although, only the $\delta B_x$ and $\delta B_z$ components were detected in vacuum, the appearance of a finite $\delta B_y$ component in the plasma indicates that the wave acquires a polarized response through wave--plasma interaction. At an exciter current of $I_p=50$ A, with an axial magnetic field of $B_z=6.2$ G and $B_H=0$ G, pronounced higher harmonics were observed extending from the fundamental propagating frequency of $150$ kHz to approximately $600$ kHz, together with several additional spectral components (Fig.~12(c)). These observations demonstrate the nonlinear spectral evolution of the propagating electromagnetic wave in magnetized laboratory plasma.

The experiments further demonstrate that the spatial propagation of the wave can be modified by the magnetic-field configuration. For $B_z=6.2$ G and $B_H=+20$ G, the wave activity is predominantly localized near the central region of the Helmholtz-field configuration, with the corresponding wave spectrum concentrated within this region (Fig.~13). When the polarity of the Helmholtz field is reversed, i.e. $B_z=6.2$ G and $B_H=-20$ G, the wave propagation is redistributed toward the region outside the Helmholtz field boundary, accompanied by a corresponding outward redistribution of the wave spectral power. The contrasting spatial behavior observed for opposite Helmholtz-field polarities provides experimental insight into the role of the magnetic-field configuration in modifying the spatial propagation and localization of whistler-frequency electromagnetic waves. Furthermore, as shown by Whitmer et al.~\cite{49} The strength and characteristics of higher harmonics depend on the axial magnetic field, so a Helmholtz magnetic field could be useful for further investigating how higher harmonics change as the DC magnetic field varies.

In order to obtain the benefits associated with the excitation system, dedicated control experiments were carried out under identical conditions by reducing the loop-antenna excitation current from the nominal value of $50$ A to $10$ A and $6$ A. At these lower excitation amplitudes, the measured spectrum was dominated by the excitation component $50$ kHz and its second harmonic at $100$ kHz, while the pronounced higher-frequency harmonic structure observed at $I_p=50$ A was absent (Fig.~12(a,b)).

Propagation delay measurements and cross-correlation analysis demonstrated strong spatial coherence of the propagating wave packet. The experimentally determined dispersion relationship shows reasonable agreement with the analytical cold-plasma oblique whistler dispersion relationship evaluated using independently measured plasma parameters. The measured polarization and ellipticity over the investigated frequency range, and the experimentally determined wave-normal angle of approximately $48^\circ$ are also consistent with oblique whistler propagation.

An important feature of the experiment is that the dominant propagating mode in the plasma is centered at $150$ kHz, Although the loop-antenna current contains a strong 50-kHz component, the dominant propagating electromagnetic whistler mode identified in the plasma is centered at 150 kHz; the 150-kHz component is therefore taken as the fundamental propagating whistler mode for the subsequent harmonic analysis. The bicoherence analysis further indicates significant nonlinear coupling among the observed spectral components, providing evidence for wave--wave interactions associated with the observed frequency generation and spectral redistribution. The spatial evolution of the harmonic spectrum over a radial distance of $50$ cm, together with the experimentally determined spectral power conversion efficiencies up to the sixth harmonic, further demonstrates the nonlinear evolution and spectral energy redistribution of the propagating whistler wave.
For the present experimental conditions ($B_H=20$ G and $B_z=6.2$ G), the lower-hybrid frequency lies below the $150$--$900$ kHz frequency range of the higher-frequency components investigated here. Moreover, the predominantly electrostatic and strongly oblique character of the lower-hybrid mode differs from the finite electromagnetic fluctuations with finite $k_\parallel$ observed in the present experiments; hence, no direct role of the lower-hybrid resonance is identified in the observed wave propagation or harmonic-generation process.

Although the present measurements provide experimental evidence for higher-harmonic generation, frequency mixing, and nonlinear spectral evolution of whistler-mode waves, the precise physical mechanism responsible for the generation of the observed higher-frequency components cannot yet be uniquely identified from the present measurements. The observed spectral evolution is consistent with nonlinear wave--plasma interactions, in which higher-order nonlinear current densities arising from the plasma fluid response can act as sources of harmonic generation. In addition, Vasko \textit{et al.}~\cite{57} demonstrated that substantial harmonic power can arise from nonlinear electrostatic steepening of finite-amplitude whistler waves through coupling with electron-acoustic modes in a two-temperature plasma. Therefore, in addition to conventional nonlinear wave--wave interactions, electrostatic steepening may also contribute to the spectral features observed in the present experiment. Further measurements of the associated electrostatic fluctuations and nonlinear mode coupling would be required to distinguish quantitatively between these mechanisms.

\section{Acknowledgment}
The authors sincerely thank Vishal Sharma for his help during the campaign phase and Mr Dashrath for his help during the diagnostic installations.

\bibliographystyle{unsrt}
\bibliography{ref}

\end{document}